\documentclass[10pt,a4paper]{article}
\usepackage{amssymb}
\newtheorem{thm}{Theorem}
\newtheorem{defn}{Definition}
\newtheorem{cor}{Corollary}
\newtheorem{lem}{Lemma}

\newtheorem{prop}{Proposition}

\newtheorem{rk}{Remark}

\usepackage{amssymb}

\def \C{{\! \rm \
I \!\!\!C}}

\def \Q {{\! \rm \ I\!\!\! Q}}
\def \A {{\! \rm \ I \!\! A}}
\def \E {{\! \rm \ I \!E}}
\def \N {{\! \rm \ I \!N}}
\def \M {{\! \rm \ I \!\!\! M}}
\def \Z {{\! \rm Z\! \!Z}}

\def\otherterm#1{{\it#1}}
\def \e {{\epsilon}}
\newcommand{\tr} {\hbox{tr}}

\def \l {{\lambda}}
\def \tr {{\rm tr}}

\begin{document}

\title{\bf   Weighted trace cochains; a geometric setup for anomalies}
\author{ Sylvie
PAYCHA }
\maketitle

\section*{Abstract} 
We extend formulae which measure discrepancies for regularized traces 
on classical pseudodifferential operators \cite{MN}, \cite{CDMP}, \cite{PR}
to regularized trace cochains, regularized traces corresponding to $0$-regularized
trace cochains. This extension from $0$-cochains to $n$-cochains is appropriate
to handle simultaneously algebraic and geometric discrepancies/anomalies due
to the presence of a weight.   {\it Algebraic anomalies} are  Hochschild
coboundaries of regularized trace cochains on a fixed algebra of 
pseudodifferential operators  weighted by a fixed classical 
pseudodifferential operator $Q$ with positive order   and positive scalar 
leading symbol. In contrast, {\it geometric anomalies} arise when considering 
 families of algebras of  pseudodifferential operators associated with  a
smooth fibration of manifolds. They correspond to   covariant derivatives (and
possibly their curvature) of smooth families of regularized trace cochains, the
weight being here an elliptic operator valued form on the base manifold. Both
types of discrepancies can be expressed as finite linear combinations  of
Wodzicki residues. We apply the formulae obtained in the family setting  to
build Chern-Weil type weighted trace cochains on one hand, and on the other hand to show that
choosing the curvature of a Bismut-Quillen type super connection as a weight,
provides covariantly closed  weighted trace cochains in which case the geometric discrepancies  vanish.

\section*{Introduction}  Linear forms  $A\mapsto 
{\rm tr} \left(Ae^{-\e Q}\right)$  on the algebra $Cl(M, E)$ of classical
pseudodifferential operators acting on smooth sections of a hermitian vector
bundle $E$ based on a closed Riemannian  manifold $M$ where $Q$ is a classical
pseudodifferential operator with positive scalar  leading symbol and positive
order, naturally generalise to  multilinear forms, namely  JLO   cochains
\cite{JLO} (see also  \cite{P} in relation to anomalies):  \begin{eqnarray*}
\tilde \chi_{n, Q}(\e)(A_0, \cdots, A_n)&=&  \int_{\Delta_n}du_0\cdots du_n \,
{\rm tr}\left( A_0 e ^{-\e\cdot u_0 Q}  A_1  e^{-\e\cdot u_1 Q} \cdots
\right.\\ &{}& \left. A_{n-1} e^{-\e \cdot u_{n-1} Q} A_n e^{-\e \cdot u_n
Q}\right), \end{eqnarray*} since $\tilde \chi_{0, Q}(\e)(A)= {\rm tr}(Ae^{-\e
Q})$.  Here $\Delta_n:= \{(u_0, \cdots, u_n)\in [0, 1]^{n+1}, \sum_{i=0}^n
u_i=1\}$.

Recall that \footnote{This holds provided $Q$ is invertible, otherwise we turn
it into an invertible operator adding the orthogonal projeciton onto its
kernel} the Mellin transform of $t\mapsto {\rm tr} \left(A e^{-t Q}\right)$ 
defines a  meromorphic function $z\mapsto {\rm TR} (A Q^{-z})$ (here TR stands
for the canonical trace introduced in \cite{KV}) with simple poles  and 
complex residue given by $\frac{1}{q} {\rm res} (A)$ where $q$ is the order of
$Q$ and res$(A)$ the Wodzicki residue of $A$ \cite{Wo}. Similarly, we show
that  the Mellin transform of $t \mapsto  \tilde \chi_{n, Q}(t)(A_0, \cdots,
A_n)$ yields meromorphic functions $z\mapsto \bar \chi_{n, Q}(z)(A_0, \cdots,
A_n)$   with simple poles and  complex residue given by $\frac{1}{q}{\rm
res}(A_0\cdots A_n)$. \\ On the other hand, the finite part of $z\mapsto {\rm
TR} (A Q^{-z})$ defines a very useful linear map $A\mapsto {\rm tr}^Q(A)$
which we refer to as {\it $Q$-weighted trace of $A$}  \cite{P}, \cite{CDMP},
\cite{MN}. In the same way, the finite part of  $z\mapsto \bar \chi_{n,
Q}(z)(A_0, \cdots, A_n)$ defines  {\it $Q$-weighted trace  cochains}
$\chi_n^Q( A_0, \cdots, A_n)$  which yield cyclic cochains that differ from 
the weighted trace ${\rm tr}^Q\left(A_0 A_1 \cdots A_n \right)$ of the product
of the $A_i$'s (which is not cyclic) by a finite linear combination of
Wodzicki residues (Proposition \ref{prop:barchi}): \begin{eqnarray} &{}&
\chi_n^Q(A_0, \cdots, A_{n}) ={\rm tr}^Q\left(A_0 A_1 \cdots A_n
\right)\nonumber\\ &+&\frac{1}{q} \sum_{\vert k\vert=1}^{[\vert a\vert] +{\rm
dim} M} \frac{(-1)^{\vert k\vert} (\vert k\vert-1)!}{ (k+1)!} \, {\rm res}
\left(A_0 A_1^{(k_1)} \cdots A_n^{(k_n)} Q^{-\vert k\vert}\right).
\end{eqnarray} Here $A^{(j)}:= {\rm ad}_Q^j(A)$,  $\vert a\vert:= a_0+\cdots
+a_n$ is the order of the product $A_0\cdots A_n$, $[\vert a\vert ]$ its
integer part and  for any multiindex $k=(k_1, \cdots, k_n)$, we set $\vert
k\vert = k_1+\cdots +k_n$, $(k+1)!=( k_1+1)!\cdots( k_n+1)!$. \\ \\ The
presence of a weight $Q$ leads to discrepancies which are often responsible
for the occurence of (local) anomalous phenomena in physics and infinite
dimensional geometry \cite{CDMP}, \cite{CDP}, \cite{PR}.  We consider two 
types of discrepancies, algebraic and geometric ones; the first type arises 
as  Hochschild coboundaries of weighted trace cochains, the second type as 
covariant derivatives (and possibly the corresponding curvature) of families
of weighted trace cochains. Working with   families of cochains offers a
natural geometric setting that  brings together  these two types of anomalies
in a common framework.\\  We express the Hochschild coboundary of a weighted
trace $2p$-cochain as a finite linear combination of a finite number of
Wodzicki residues involving powers of the weight $Q$ (Theorem
\ref{thm:bbarchi}): \begin{eqnarray}\label{eq:1} &{}& b\,  \chi_{2p}^Q(A_0,
\cdots, A_{2p+1}) = \frac{1}{q}\sum_{\vert k\vert=0}^{ [\vert a\vert] +{\rm
dim}\, M-1}\frac{(-1)^{\vert k\vert}\, \vert k\vert!} {(k+1)!} \sum_{j=0}^{p} 
 {\rm res} \left( A_0 \,A_1^{(k_1)}\cdots\right. \nonumber\\ &{}&\left.\cdots
A_{2j}^{(k_{2j})}\, A_{2j+1}^{(k_{2j+1}+1)}\, A_{2j+2}^{(k_{2j+2})} \,   \cdots A_{2p+1}^{(k_{2p+1})} Q^{-\vert
k\vert-1}\right). \end{eqnarray} When $p=0$, this yields  $$\left(b\, {\rm
tr}^Q\right)(A, B) = \frac{1}{q}\sum_{\vert k\vert=0}^{ [ a+b] +{\rm dim}\,
M-1}\frac{(-1)^{ k}} {k+1} {\rm res} \left( A B^{(k+1)}\, Q^{- k-1}\right),$$ 
where $a$ is the order of $A$, $b$ that of $B$.  Equation (\ref{eq:1}), which
is very close in spirit to formulae derived in \cite{H}, \cite{CM} (see
Appendix A for some analogies \footnote{The essential difference lies in the
fact that the weighted trace forms are cyclic and hence generally not $(b, B)$
closed, in contrast to the Chern character}to compute character cocycles,
shows the expected locality of the algebraic anomaly since the Wodzicki
residue has an explicit local description in terms of the (positively)
homogeneous symbol $\sigma_{-{\rm dim} M}$  of order $-{\rm dim} \,M$ of the
operator: $${\rm res}(A)= \frac{1}{(2\pi)^{{\rm dim}\, M}}\int_{S^*M} dx\,
d_S\xi\, {\rm tr}_x \left( \sigma_{-{\rm dim \,M}} (A)(x, \xi)\right)$$ where $S^*M$ denotes the unit cotangent sphere and $d_S\xi$ the canonical volume measure on it.  It
also follows from this formula that  $Q$-weighted trace $2p$- cochains yield
$2p$-cocycles on the algebra  $Cl_{\leq -\frac{{\rm dim M}}{(2p+2) }} (M, E)$
which (strictly) includes the algebra  $Cl_{<- \frac{{\rm dim M}}{(2p+2) }}
(M, E)$ of  classical pseudodifferential operators that lie in the Schatten
class ${\cal I}_{2p+2}\left( L^2(M, E)\right)$. Here $L^2(M, E)$ is the
$L^2$-completion of  $C^\infty(M, E)$ w.r. to the hermitian metric on $E$ and
the Riemannian metric on $M$.  \\ Weighted trace cochains vary with the
weight; if $\Q:x\to Q_x\in Cl(M, E)$ is a smooth family of weights
parametrized by a smooth manifold $X$ then (Theorem \ref{thm:dbarchi}):
\begin{eqnarray}\label{eq:2} &{}&  \left(d \,\chi_n^\Q\, \right)   (A_0,
\cdots, A_{n} )\nonumber\\ &=&  \frac{1}{q}\cdot  \sum_{\vert
k\vert=0}^{[\vert a\vert] +{\rm dim }M}  \frac{(-1)^{\vert k\vert+1} k!}{
(k+1)!}\sum_{j=1}^{n+1 } \, {\rm res} \left( A_0 \, A_1^{(k_1)}\,\cdots\right.
\nonumber \\ &{}&\cdot \left. A_{j-1}^{(k_{j-1})} \,   \left(d\,
\Q\right)^{(k_j)} A_{j}^{(k_{j+1})}    \cdots\,  A_n^{(k_{n+1})} \Q^{-\vert k\vert-1}\right)
\end{eqnarray} For $n=0$, this gives: $$\left(d \,{\rm tr}^\Q\right)   (A ) =
\frac{1}{q}\cdot  \sum_{ k=0}^{[ a] +{\rm dim }\, M}  \frac{ (-1)^{k+1}}{ k+1}
{\rm res} \left( A \,   \left(d\, \Q\right)^{(k)} \Q^{- k-1}\right).$$ In the
family setup, to a fibration $\pi: \M\to X$ of closed Riemannian manifolds
$\{M_x, \, x\in X\}$ modelled on $M$ and   based on a smooth manifold $X$
together with  a hermitian vector bundle $\E\to \M$, we can associate a 
smooth fibration  of algebras $Cl(\M, \E)$  modelled on $Cl(M,  E)$ with fibre
over $x\in X$ given by  $Cl(M_x, E_{\vert_{M_x}})$. Given  a smooth family of
weights $\Q\in \Omega^{even}(X, Cl(\M, \E))$, we define  corresponding smooth
families  of $\Q$-weighted trace cochains  $(\alpha_0, \cdots,
\alpha_n)\mapsto \chi_{2p}^\Q(\alpha_0, \cdots, \alpha_{n})$ on
$\Omega\left(X, Cl(\M, \E)\right)$. When  the fibration $Cl(\M, \E)$ is
equipped with a connection     $\nabla$  such that locally, $\nabla \alpha =
d\alpha  + ad_{\theta}$ where $\theta$  is a local $ Cl(M, E)$-valued on form,
we express the covariant derivative  of a weighted trace $2p$-cochain as a
linear combination of a finite number of Wodzicki residues involving powers of
the weight $\Q$ (Theorem \ref{thm:nablachiQ}) \footnote{Here again, we are
assuming that $\Q$ is invertible, otherwise, provided its kernel 
defines a vector bundle, we can turn it into an invertible operator adding the
orthogonal projection onto its kernel}: \begin{eqnarray}\label{eq:3} &{}& 
\left(\nabla \,\chi_n^\Q\right)   (\alpha_0, \cdots, \alpha_{n} )\nonumber\\
&=&  \frac{1}{q}\cdot \sum_{\vert k\vert=0}^{[\vert a\vert] +{\rm dim }M}
\,\frac{k!}{ (k+1)!}\, \sum_{j=1}^{n+1 } (-1)^{\vert \alpha_0\vert +\cdots +
\vert \alpha_{j-1}\vert+\vert  k\vert +1}  {\rm res} \left( \alpha_0 \wedge
\alpha_1^{(k_1)}\wedge \cdots\right.\nonumber\\ &{}& \left. \wedge \,  
\alpha_{j-1}^{(k_{j-1})} \wedge   \left(\nabla^{End}\Q\right)^{(k_j)}\wedge \alpha_{j}^{(k_{j+1})}\wedge 
\cdots\wedge  \alpha_n^{(k_{n+1})}\wedge \Q^{-\vert k\vert-1}\right). \end{eqnarray}
For $n=0$ this yields $$  \left(\nabla \,{\rm tr}^Q\right)   (\alpha) =
\frac{1}{q}\cdot \sum_{ k=0}^{[ a] +{\rm dim }M} \, \frac{(-1)^{\vert
\alpha\vert +   k +1}}{k+1}  {\rm res} \left( \alpha\wedge   
\left(\nabla^{End}\Q\right)^{(k)} \wedge \Q^{- k-1}\right).  $$ Equation
(\ref{eq:3}) shows that geometric discrepancies  are also local in as far as
they can be written in terms of a finite number of Wodzicki residues, thus
generalizing observations already made previously \cite{PR} on the locality of
obstructions of the type $\nabla {\rm tr}^\Q= \nabla \chi_0^\Q$.\\   
Chern-Weil type $\Q$-weighted trace cochains $\chi_n^\Q(f_0(\Omega), \cdots,
f_n(\Omega))$, where the $f_i$'s are polynomial functions and $\Omega=
\nabla^2$ the curvature of $\nabla$ generalize the $\Q$-weighted Chern forms
${\rm tr}^\Q(f(\Omega))= \chi_0^\Q(f(\Omega))$ discussed in \cite{PR} which
occur in \cite{F} in a disguised form, see \cite{CDMP}. The above formula
measures the obstruction to their closedness:  \begin{eqnarray} \label{eq:4}
&{}& d  \,\chi_n^\Q    (f_0 (\Omega), \cdots, f_{n}(\Omega) )\nonumber \\ &=&
\frac{1}{q}\cdot \sum_{\vert k\vert=0}^{[\vert d\vert\cdot \omega] +n} \,
\frac{k!}{(k+1)!}\, \sum_{j=1}^{n+1 } (-1)^{\vert  k\vert +1}  {\rm res}
\left( f_0(\Omega) \wedge \left(f_1(\Omega)\right)^{(k_1)}\wedge
\cdots \wedge
\left(f_{j-1}(\Omega)\right)^{(k_{j-1})}  \right.\nonumber\\ &{}& \left.\wedge  
\left(\nabla^{End}\Q\right)^{(k_j)} \wedge \left(f_{j}(\Omega)\right)^{(k_{j+1})}\wedge \cdots\wedge
\left(f_{n}(\Omega)\right)^{(k_{n+1})} \Q^{-\vert k\vert-1}\right)
\end{eqnarray} where $\vert d\vert= \sum_{i=0}^n d_i$  with  $d_i$ the degree
of the polynomial $f_i$, and  $ \omega$ is the order of the operator valued
$2$-form $\Omega$.  As expected, these obstruction vanish in the context of
families of Dirac operators, replacing $\nabla$ by a Quillen-Bismut type
superconnection $\A= \nabla+ D$ (\cite{Q}, \cite{B}, see also \cite{BGV})  and
setting $\Q= \A^2$.This  leads to  characteristic classes built  along a line 
suggested by Scott's work  \cite{Sc} and further developped  in \cite{PS}. 
\section*{Acknowledgements} I would like to thank  Matilde Marcolli for
interesting discussions which encouraged me to write this down. I am also
grateful to  the Max Planck Institute in Bonn and Matilde Marcolli for
inviting me for a two months stay during which   this article was written. 
\section{JLO cochains on weighted trace  algebras } Let  $({\cal A},\cdot)$ 
be an associative algebra over $\C$  and let  $C^n\left( {\cal A}\right)$
denote the space of continuous  $\C$-multilinear    valued forms on ${\cal
A}^{\otimes n+1}$, which  corresponds to the space of $n$-cochains on ${\cal
A}$. The Hochschild coboundary  of an  $n$-cochain $\chi_n$ is an
$n+1$-cochain  defined by: 
\begin{eqnarray*}
b\chi_n(A_0,\cdots, A_{n+1})&=& \sum_{j=0}^n (-1)^j
\chi_n(A_0, \cdots, A_j \cdot A_{j+1}, \cdots, A_{n+1})\\
&+& (-1)^{n+1}
\chi_n(A_{n+1} \cdot A_0, \cdots, A_n).
\end{eqnarray*} Since  $b^2=0$, this  defines a 
cohomology, called the  Hochschild cohomology  of ${\cal A}$. In the
following, we shall drop the explicit $\cdot$ in  the product notation writing
simply $AB$ for $A\cdot B$.  \begin{defn}   A weighted tracial algebra is a
triple  $\left({\cal A}, Q, T\right)$ where ${\cal A}$ is a topological unital
associative algebra, $T$ is a  continuous linear map on a non trivial  ideal
${\cal I}$ of ${\cal A}$ and $Q$ an element of ${\cal A} $ such that
\begin{itemize} \item $T$ is a  trace on ${\cal I}$, \item   the equation
$\frac{d}{dt} U_t= Q\,  U_t$ with initial condition $ U_0= 1_{\cal A}$ has a
unique solution $U_t=e^{-t Q}\in {\cal A},\, t\geq 0$,  \item   $e^{-t Q}\in
{\cal I}$ for any $0<t$.   \end{itemize}   $Q$ is called the weight. \\ Given
a weighted tracial algebra $\left({\cal A}, Q, T\right)$, we define a JLO type
  cochain \cite{JLO} (see also \cite{G-BVF}): $$\chi_{ n,Q}(A_0, \cdots,
A_n):=\int_{\Delta_n}\,du_0\cdots du_n \, T\left( A_0 e^{-u_0 Q}  A_1  e^{-u_1
Q} \cdots A_{n-1}  e^{-u_{n-1} Q}A_n e^{-u_nQ}\right).$$  \end{defn}
\begin{lem}\label{lem:ad} Setting ${\rm ad}_AB:= [A, B]$ we have:  $${\rm
ad}_{A}e^{-u Q}= -u \int_0^1 dt\, e^{-u(1-t)Q} \left({\rm ad}_AQ\right) e^{-ut
Q}=  u \int_0^1 dt\, e^{-u(1-t)Q} \left({\rm ad}_{Q} A\right) e^{-ut Q}. $$
\end{lem} {\bf Proof:} (see e.g. \cite{G-BVF}) Differentiating the map
$u\mapsto [e^{-u Q}, A]$, we get $$\left(\frac{d}{du} +Q\right)[e^{-u Q}, A]=
[A,Q] e^{-u Q}.$$ Solving this equation by the usual Duhamel formula for first
order inhomogeneous linear differential equations gives $[e^{-u Q}, A]=
\int_0^u e^{-s Q} [Q, A] e^{-(u-s)Q }ds$. Substituting $s=u t$ yields the
above formula.\\ \\ As a consequence, we have
\begin{prop}\label{prop:ad} If   
$n=2k$ is even then  \begin{eqnarray*} &{}&b\, \chi_{2k, Q}(A_0, \cdots,
A_{2k+1})\\
&=&\sum_{j=0}^{k}  \chi_{2k+1, Q}\left( A_0, A_1,\cdots,A_{2j},  {\rm
ad}_{Q}A_{2j+1},A_{2j+2},   \cdots , A_{2k+1}\right) \end{eqnarray*}
If $n=2k+1$ is odd,  \begin{eqnarray*} &{}&b\, \chi_{2k+1,Q}(A_0, \cdots,
A_{2k+2})\\&= &   \sum_{j=0}^{k}  \chi_{2k+2, Q}\left( A_0,
A_1,\cdots,\cdots,A_{2j}, {\rm ad}_{Q}  A_{2j+1}, A_{2j+2}, \cdots ,
A_{2k+2}\right)\\ &- &   \chi_{2k+1, Q} \left(A_{2k+2} \cdot A_0,A_1, \cdots,
A_{2k+1}\right). \end{eqnarray*} \end{prop} {\bf Proof:} We carry out the
proof in the even case. The odd case can be derived similarly.    First
observe that  given weights $Q_0, \cdots, Q_n\in {\cal A}$, we can define the
$n$-cochain $\chi_{ n,Q_0,\cdots , Q_n}$ by:  $$\chi_{ n,Q_0,\cdots ,
Q_n}(A_0, \cdots, A_n):= T\left( A_0 e^{-Q_0} A_1  e^{-Q_1} \cdots A_{n-1} 
e^{-Q_{n-1}}  A_n e^{-Q_n}\right).$$ A straightforward computation shows
\begin{eqnarray*} &{}& \chi_{n,Q_0,\cdots , Q_n }(A_0, \cdots,A_{j-1}
A_j,\cdots, A_{n}) -\chi_{n,Q_0,\cdots , Q_n }(A_0, \cdots,A_{j}
A_{j+1},\cdots, A_{n})\\ &=& T (A_0  e^{-Q_0}\cdots A_{j-1} [  A_{j}, 
e^{-Q_{j-1}}]\,  e^{-Q_j}  A_{j+1}\cdots  A_{n}\,  e^{-Q_n}).\\
\end{eqnarray*} Setting $Q_j= u_j \, Q$ it follows that \begin{eqnarray*} &{}&
b\, \chi_{2k,Q}(A_0, \cdots, A_{2k+1})\\ &= &  - \sum_{j=0}^{k}
\int_{\Delta_{2k}}d u_0\cdots du_{2k}\, u_{2j} \,  T \left( A_0 e^{-u_0 Q} 
A_1 e^{-u_1 Q}\cdots A_{2j}\cdot \right.\\ &\cdot&\left.\left(\int_0^1
dt \,e^{-u_{2j} (1-t) Q}  [A_{2j+1}, Q]  e^{-u_{2j} t Q} \right) \cdots
A_{2k+1}e^{-u_{2k} Q}\right)\\ &= &  - \sum_{j=0}^{k} \int_{\Delta_{2k}}d
u_0\cdots du_{2k}  \int_0^{u_{2j}} du \,  T \left( A_0 e^{-u_0 Q}   
A_1e^{-u_1 Q}\cdots A_{2j} e^{(-u_{2j} + u) Q}  [A_{2j+1},Q]\right.\\
&\cdot&\left. e^{-u Q}  \cdots A_{2k+1} e^{-u_{2k} Q}\right)\\ &=& -
\sum_{j=0}^{k}\int_{\Delta_{2k+1}}d u_0\cdots du_{2k+1}    T \left( A_0
e^{-u_0 Q} \cdots A_{2j} e^{-u_{2j} Q} [A_{2j+1}, Q]\right. \\ &\cdot& \left.
e^{-u_{2j+1} Q} A_{2j+2} \cdots A_{2k+1} e^{-u_{2k+1} Q}\right) \\
&=&\sum_{j=0}^{k}  \chi_{2k+1, Q}\left( A_0, A_1,\cdots, [ Q,A_{2j+1}], 
\cdots , A_{2k+1}\right). \end{eqnarray*} \section{Weighted trace cochains on
pseudodifferential operators}  Let ${\cal A}= Cl(M, E)$ be  the algebra of
classical pseudodifferential operators (PDOs) acting on smooth sections of a
vector bundle $E$ based on a closed Riemannian manifold $M$ and let ${\cal I}$
the ideal of smoothing operators equipped with the ordinary trace $T={\rm
tr}$.  Let  $Q\in Cl(M, E)$  be an   operator with positive scalar leading
symbol and positive order $q$. Since its leading symbol is invertible, $Q$ is
elliptic.  Then $({\cal A}, Q, {\rm tr})$ is a weighted tracial algebra.  We
call the couple $\left(Cl(M, E), Q\right)$ a  {\it weighted PDO algebra}. 
Replacing $Q$ by    $Q_\e:= \e\, Q$ for some fixed $0<\e<1$ in the JLO type
cochains associated to the weighted trace algebra  $({\cal A}, Q, {\rm tr})$,
we  define: \begin{eqnarray*} \tilde \chi_{n, Q}(\e)(A_0, \cdots, A_n)&=& 
\int_{\Delta_n}du_0\cdots du_n \, {\rm tr}\left( A_0 e ^{-\e \, u_0 Q}  A_1\, 
e^{-\e \, u_1 Q} \cdots \right.\\ &{}& \left.\cdots A_{n-1} e^{-\e \, u_{n-1}
Q} A_n e^{-\e \, u_n Q}\right). \end{eqnarray*} The following lemma is an
immediate  consequence of Proposition \ref{prop:ad} applied to \hfill
\break \noindent$Q_t=t Q$: \begin{lem}\label{lem:btildechi} Given a weighted  pseudodifferential
algebra $\left(Cl(M, E), Q\right)$, for any  pseudodifferential operators $A_0,\cdots, A_{2p+1}$ in $Cl(M,
E)$  \begin{eqnarray*} &{}& b\,\tilde  \chi_{2p, Q}(t)(A_0, \cdots,
A_{2p+1})\\&=& t \, \sum_{j=0}^{p}  \chi_{2p+1, Q}(t) \left( A_0,
A_1,\cdots,A_{2j},  {\rm ad}_Q A_{2j+1} ,A_{2j+2},   \cdots ,
A_{2p+1}\right).\\ \end{eqnarray*} \end{lem} \begin{defn}  Whenever $Q$ is
invertible, the  Mellin transform of $\tilde \chi_{n, Q}$ is given by:
\begin{eqnarray*}&{}& \bar \chi_{2k, Q}(z)(A_0, \cdots, A_{n})\\
&:=&
\frac{1}{\Gamma(z)}  \int_0^\infty  t^{z-1} \tilde \chi_{2k, Q}(t)(A_0,
\cdots, A_{n})\, dt\\ &=& \frac{1}{\Gamma(z)} \int_0^\infty t^{z-1} dt\, 
\int_{\Delta_n}du_0\cdots du_n \, {\rm tr}\left( A_0 e ^{-t\cdot u_0 Q} A_1 
e^{-t\cdot u_1 Q} \cdots \right.\\ &{}& \left. A_{n-1} e^{-t \cdot u_{n-1} Q}
A_n e^{-t \cdot u_n Q}\right). \end{eqnarray*} \end{defn}  \begin{rk} When $Q$
is not invertible, since it has a finite dimensional kernel being an elliptic
operator on a closed manifold, assuming that $E$ comes equipped with a
hermitian structure, we can take instead: $$\tilde Q:= Q+\pi_{\rm Ker Q}$$
where $\pi_{\rm Ker Q}$ is the orthogonal projection onto the kernel of $Q$.
In the following, we therefore assume $Q$ is invertible unless otherwise
specified.  \end{rk} To state the next result, we need some notations.
\begin{defn}  For $j\in \N$ and $A$ in the weighted PDO algebra $ \left(Cl(M,
E), Q\right)$, we set as in the introduction: $$A_Q^{(0)}:= {\rm ad}_Q^j(A),
\quad {\rm where} \quad  {\rm ad}_Q(B)=[Q, B],$$  so that $ A_Q^{(0)}=I, \quad
 A_Q^{(j+1)}= {\rm ad}_Q(A^{(j)})=[Q, A^{(j)}].$ \end{defn} We shall often
drop the subscript $Q$, writing $A^{(j)}$ instead of $A_Q^{(j)}$. It is useful
to notice that since  $Q$ has scalar leading symbol then $A^{(j)}$ has order
$a+j(q-1)$ if $a$ is the order of $A$ where $a$ the order of $a$. 
\begin{prop}\label{prop:barchi} Given any $A_0, \cdots, A_n\in Cl(M,E)$, the
map $z\mapsto \bar \chi_{n, Q}(z)(A_0, \cdots, A_{n})$ is meromorphic with
simple poles. Its complex residue at $z=0$ is given by: $$ {\rm Res}_{z=0} 
\bar \chi_{n, Q}(z)(A_0, \cdots, A_{n})=\frac{1}{q}\, {\rm res} \left(A_0 A_1
\cdots A_n \right).$$ Its finite part  at $z=0$ called the {\bf $Q$-weighted
$n$-trace cochain} is given by: $$ \chi_0^Q(A_0):= {\rm tr}^Q(A_0):= {\rm
fp}_{z\to 0}\bar \chi_{0, Q}(z)(A_0)$$ when $n=0$ and for any $n\in \N$ by 
\begin{eqnarray*}   \chi_n^Q(A_0, \cdots, A_{n}) &:=& {\rm fp}_{z=0}  \bar
\chi_{n, Q}(z)(A_0, \cdots, A_{n})\\ &= &{\rm tr}^Q\left(A_0 A_1 \cdots
A_n\right)\\
&+&\frac{1}{q} \sum_{\vert k\vert=1}^{[\vert a\vert] +{\rm dim } M}
\frac{(-1)^{\vert k\vert}(\vert k\vert-1)!}{( k+1)!}\, {\rm res} \left(A_0
A_1^{(k_1)} \cdots A_n^{(k_n)} Q^{-\vert k\vert}\right)\\ \end{eqnarray*}
where $q>0$ is the order of $Q$, $\vert a\vert:= a_0+\cdots +a_n$ the order of
the product $A_0\cdots A_n$, $[\vert a\vert ]$ its integer part and where for
any multiindex $k=(k_1, \cdots, k_n)$ we have set $\vert k\vert = k_1+\cdots
+k_n$, $(k+1)!= (k_1+1)!\cdots (k_n+1)!$.  \end{prop} \begin{rk} This
proposition  shows  that $\chi_n^Q(A_0, \cdots, A_{n})$ and ${\rm
tr}^Q\left(A_0 A_1 \cdots A_n\right)$, which are two different ways of
defining a "regularized trace" of the product $A_0\cdots A_n$, differ by a
finite linear combination of Wodzicki residues. The first part of the
proposition says that on the level of residues, it does not make any
difference what regularization one considers.  \end{rk} {\bf Proof:}  First
observe that (see Appendix B) for each $J\in \N$,  there exist positive integers  $N_1,
N_2,\cdots, N_n$  such that  \begin{eqnarray} &{}&\label{eq:basicformula} \tilde
\chi_{n, Q}(t)\left(A_0 , A_1, \cdots, A_n\right)\nonumber\\
&=& \sum_{k_1=0}^{N_1-1}\cdots
\sum_{k_n=0}^{N_n-1} \frac{(-1)^{\vert k\vert} t^{\vert k\vert}}{ (k+1)!} 
{\rm tr} \left( A_0 \,A_1^{(k_1)} \cdots A_n^{(k_n)} e^{-tQ}\right)+o(t^J).
\end{eqnarray}  Hence, for each ${\rm Re}( z)\leq A$, we can choose the
$N_i$'s large enough so that the rest term vanishes in the following
computations: \begin{eqnarray}\label{eq:dbarchinq} &{}&\bar \chi_{n, Q}(z)(A_0,
\cdots, A_{n})\nonumber\\&:=& \frac{1}{\Gamma(z)}  \int_0^\infty  t^{z-1}
\tilde \chi_{n, Q}(t)(A_0, \cdots, A_{2k+1})dt\nonumber \\ &=&
\frac{1}{\Gamma(z)} \sum_{k_1=0}^{N_1-1}\cdots \sum_{k_n=0}^{N_n-1}
\frac{(-1)^{\vert k\vert} \,t^{\vert k\vert}}{( k+1)!}   \int_0^\infty dt\,
t^{z-1}\,{\rm tr} \left( A_0 \,A_1^{(k_1)} \cdots A_n^{(k_n)}
e^{-tQ}\right)\nonumber\\ &=& \frac{1}{\Gamma(z)} \sum_{k_1=0}^{\infty}\cdots
\sum_{k_n=0}^{\infty} \frac{(-1)^{\vert k\vert} }{ (k+1)!}  \int_0^\infty
dt\,t^{z+\vert k\vert-1}\, {\rm tr} \left( A_0 \,A_1^{(k_1)} \cdots
A_n^{(k_n)} e^{-tQ}\right)\nonumber\\ &=& \sum_{k_1=0}^{\infty}\cdots
\sum_{k_n=0}^{\infty} \frac{(-1)^{\vert k\vert}}{ (k+1)!}  
\frac{\Gamma(z+\vert k\vert)}{\Gamma(z)}  \, {\rm TR} \left( A_0 \,A_1^{(k_1)}
\cdots A_n^{(k_n)} Q^{-\vert k\vert -z}\right),  \end{eqnarray} where TR
stands for the canonical trace on all non integer order pseudodifferential
operators.  If $\vert k\vert=0$ then  $\frac{\Gamma(z+\vert
k\vert)}{\Gamma(z)}=1$ and  this sum reduces to  $$ {\rm TR} \left( A_0
\,A_1^{(k_1)} \cdots A_n^{(k_n)} Q^{ -z}\right)$$ which is meromorphic with
simple pole at $0$ given by  $${\rm Res}_{z=0}{\rm TR} \left( A_0 \,A_1 \cdots
A_n Q^{-z}\right) = \frac{1}{q} {\rm res} \left( A_0 \,A_1 \cdots
A_n\right).$$ If $\vert k \vert\neq 0$ then   $\frac{\Gamma(z+\vert
k\vert)}{\Gamma(z)} \sim_0(\vert k\vert-1)!z $. \\
Since the  the map  $z\mapsto
{\rm TR} \left( A_0 \,A_1^{(k_1)} \cdots A_n^{(k_n)} Q^{-\vert k\vert
-z}\right)$ is also meromorphic with a simple pole at zero,  it follows that 
when $\vert k\vert \neq 0$, the expression \hfill\break \noindent $\frac{\Gamma(z+\vert
k\vert)}{\Gamma(z)}   \, {\rm TR} \left( A_0 \,A_1^{(k_1)} \cdots A_n^{(k_n)}
Q^{-\vert k\vert -z}\right)$ converges when $z$ tends to zero to $(\vert
k\vert-1)!$ times  the  complex residue at zero of ${\rm TR}\left( A_0 \,A_1
\cdots A_n\, Q^{\vert k\vert -z}\right)$  given by: $${\rm Res}_{z=0}{\rm TR}
\left( A_0 \,A_1^{(k_1)} \cdots A_n^{(k_n)} Q^{-\vert k\vert -z}\right) = 
\frac{1}{q}{\rm res} \left( A_0 \,A_1^{(k_1)} \cdots A_n^{(k_n)} Q^{-\vert
k\vert}\right),$$ and finite part ${\rm tr}^Q(A_0\cdots A_n)$.\\  Let us now
check that the infinite sums are in fact finite. Since $Q$ has scalar leading
symbol, the operators $A_0 A_1^{(k_1)} \cdots A_n^{(k_n)} Q^{-\vert k\vert}$
have order  $\sum_{i=1}^n a_i + \sum_{i=1}^n k_i (q-1)-\vert k\vert q=
\sum_{i=1}^n a_i - \sum_{i=1}^n k_i = \vert a\vert-\vert k\vert $  which
decreases as the indices $k_i$ increase. Since the Wodzicki residue vanishes
on operators of order smaller than $-{\rm dim} M$,  only the terms in the  the
sum on the r.h.s such that $\vert k\vert \leq \vert a\vert +{\rm dim} M$ 
survive so that the sum is indeed finite and stops at $[\vert a\vert] +{\rm
dim} M$. This ends the proof of the proposition. 
\begin{thm}\label{thm:bbarchi} Given any $A_0, \cdots, A_{2p+1}\in Cl(M,E)$,
the coboundary  $$z\mapsto b \,\bar \chi_{2p, Q}(z)(A_0, \cdots, A_{2p+1})$$ is
holomorphic at zero. Its value  at $z=0$ is given by: \begin{eqnarray*}  b\, 
\chi_{2p}^Q(A_0, \cdots, A_{2p+1})&=& \frac{1}{q} \sum_{\vert k\vert=0}^{
[\vert a\vert] +{\rm dim}\, M-1}(-1)^{\vert k\vert}\, \frac{ \vert
k\vert!}{(k+1)!} \sum_{j=0}^{p}   {\rm res} \left( A_0
\,A_1^{(k_1)}\cdots\right.\\ &{}& \left.\cdots A_{2j}^{(k_{2j})}\,  A_{2j+1}^{(k_{2j+1}+1)} \,  A_{2j+2}^{(k_{2j+2})}
\cdots A_{2p+1}^{(k_{2p+1})} Q^{-\vert k\vert-1}\right) \end{eqnarray*} where 
$\vert k\vert:= k_1+\cdots+ k_{2p+1}$, $(k+1)!= (k_0+1)!\cdots (k_n+1)!$ and
$[\vert a\vert]$ is the integer part of $\vert a\vert:= a_0+\cdots+a_{2p+1}$
which corresponds to  the order of the product $A_0\cdots A_{2p+1}$. \end{thm}
{\bf Proof:} By Lemma \ref{lem:btildechi} we have 
 \begin{eqnarray*} &{}&b\,\tilde  \chi_{2p, Q}(t)(A_0,
\cdots, A_{2p+1})\\
&=&t\sum_{j=0}^{p}  \chi_{2p+1, Q}(t) \left( A_0,
A_1,\cdots,A_{2j},  [Q,A_{2j+1}],A_{2j+2},   \cdots ,
A_{2p+1}\right).
\end{eqnarray*}
Using the results of Appendix B this yields:
 \begin{eqnarray*}
&{}& b\, \bar \chi_{2k, Q}(z)(A_0, \cdots, A_{2p+1})\\ &=& \frac{1}{\Gamma(z)}
 \int_0^\infty  dt\, t^{z-1}\, b\,  \tilde \chi_{2k, Q}(t)(A_0, \cdots,
A_{2p+1})dt\\ &= &\frac{1}{\Gamma(z)}   \sum_{j=0}^{p}   
\sum_{k_1=0}^{\infty}\cdots \sum_{k_{2p+1}=0}^{\infty}  \frac{(-1)^{\vert
k\vert} }{ (k+1)!} \int_0^\infty dt\,t^{z+\vert k\vert}\, {\rm tr} \left( A_0
\,A_1^{(k_1)} \cdots [Q, A_{2j+1}]^{(k_{2j+1})}\right.\\
&{}&\left. A_{2j+2}^{(k_{2j+2})} \cdots  
A_{2p+1}^{(k_{2p+1})} e^{-tQ}\right)\\ &= &   \sum_{j=0}^{p}   
\sum_{k_1=0}^{\infty}\cdots \sum_{k_{2p+1}=0}^{\infty}  \frac{(-1)^{\vert
k\vert+1}}{ (k+1)!} \frac{\Gamma( z+\vert k\vert+1)}{\Gamma(z)}  {\rm TR}
\left( A_0 \,A_1^{(k_1)} \cdots A_{2j+1}^{(k_{2j+1}+1)}\right.\\
&{}&\left. A_{2j+2}^{(k_{2j+2})} \cdots   A_{2p+1}^{(k_{2p+1})} Q^{ - (
z+\vert k\vert+1)} \right). \end{eqnarray*} Since $ \frac{\Gamma( z+\vert
k\vert+1)}{\Gamma(z)}= \vert k\vert!\cdot z$ for any multiindex $k$ (including
the case $\vert k\vert=0$) and since $ {\rm TR} \left( A_0 \,A_1^{(k_1)}
\cdots  A_{2j}^{(k_{2j})}\, A_{2j+1}^{(k_{2j+1}+1)} \cdots   A_{2p+1}^{(k_{2p+1})} Q^{ - ( z+\vert
k\vert+1)} \right)$ has a simple pole at zero given by $\frac{1}{q}   {\rm
res} \left( A_0 \,A_1^{(k_1)} \cdots A_{2j+1}^{(k_{2j+1}+1)} \cdots  
A_{2p+1}^{(k_{2p+1})} Q^{ - ( \vert k\vert+1)} \right)$, it follows that \hfill \break \noindent $b\,
\bar \chi_{2k, Q}(z)(A_0, \cdots, A_{2p+1})$ converges to 
 \begin{eqnarray*}
\lim_{z\to 0}b\, \bar \chi_{2k, Q}(z)(A_0, \cdots, A_{2p+1}) & = &\frac{1}{q}
\sum_{k=0}^\infty (-1)^{\vert k\vert}\frac{k!}{(k+1)!} \sum_{j=0}^{p}   {\rm
res} \left( A_0 \,A_1^{(k_1)}\cdots A_{2j}^{(k_{2j})}\right.\\ &{}&\left. \cdot
A_{2j+1}^{(k_{2j+1}+1)}\, A_{2j+2}^{(k_{2j+2})}  \cdots A_{2p+1}^{(k_{2p+1})} Q^{-\vert
k\vert-1}\right). \end{eqnarray*} To finish, let us check that the sum on the
r.h.s  is finite. Because $Q$ was chosen with scalar leading symbol,  only the
terms such that  $\sum_{i=0}^{2p+1} a_i +q-1 + \sum_{i=0}^{2p+1}
k_i(q-1)-q\vert k\vert -q= \vert a\vert- \vert k\vert-1\, \geq -{\rm dim } M$
survive i.e. such that $\vert k\vert\leq \vert a\vert +{\rm dim } M-1$, which
ends the proof.  \begin{rk}  For $p=0$, the above proposition yields: 
\begin{eqnarray*}   b\,{\rm tr}^Q(A,B)&= & b\, \chi_0^Q(A,B)\\ &=&\frac{1}{q} 
\sum_{k=0}^{ [a+b] +{\rm dim} M-1} \frac{(-1)^{ k}}{k+1} \   {\rm res} \left(A 
B^{(k+1)} \, Q^{-k-1}\right) \\ \end{eqnarray*} where $a$ is the order of $A$,
$b$ the order of $B$. This formula which seems a priori more complicated than
the more compact formula \cite{MN}, \cite{CDMP}: $$b \, {\rm tr}^Q(A, B)=
-\frac{1}{q} {\rm res} (A \,[ \log Q, B])$$  bares over the latter formula the
advantage  that it does not require computing the symbol of the logarithm of
$Q$. \end{rk} \begin{cor}\label{cor:bbarchi} Given any $A_0, \cdots,
A_{2p+1}\in Cl(M,E)$ with orders $a_0, \cdots, a_{2p+1}$ such that $[\vert
a\vert]\leq -{\rm dim }M$ then  $$  b\,   \chi_{2p}^Q(A_0, \cdots,
A_{2p+1})=0.$$ In particular $Q$-weighted trace $2p$-cochains yield Hochschild
cocycles  on the subalgebra $Cl_{\leq -\frac{{\rm dim}M }{2p+2 }}(M, E)$.
\end{cor} {\bf Proof:} This follows from the fact that the residue terms in
Theorem \ref{thm:bbarchi} then vanish because a Wodzicki residue vanishes on
pseudodifferential operators of order smaller than  $-{\rm dim}\, M$. 
\begin{rk} Notice that, with the notations of the introduction, the inclusion 
$Cl(M, E)\cap {\cal I}_{2p+2}(L^2(M, E))\subset   Cl_{\leq -\frac{{\rm dim}M
}{2p+2 }}(M, E)$ is strict. \end{rk} \section{Varying the weight } Let us  now
consider a fixed associative unital topological algebra ${\cal A}$ together
with an ideal ${\cal I}$ equipped with a trace $T$ and a   smooth family of
weights  $\Q: x\mapsto Q_x\in {\cal A}$ so  that for any $x\in X$, the triple
$\left({\cal A}, Q_x, T\right)$ is a weighted tracial algebra. We set for
$\chi_n\in C^n({\cal A})$: \begin{eqnarray*} \left(d\,  \chi_n\right)(A_0,
\cdots, A_n)&:=& d\left( \chi_n(A_0, \cdots, A_n)\right)\\ &-&\sum_{j=0}^{n} 
\chi_n (A_0, \cdots, d \, A_j, A_{j+1}, \cdots, A_n) \end{eqnarray*}  where
$A_0, \cdots, A_n \in C^\infty(X,Cl(M, E))$. This defines a $C^n({\cal
A})$-valued $1$-form on $X$.   \begin{lem}\label{lem:dchi} $$d\,  e^{-t\Q}= -t
\int_0^1 du e^{-t(1-u)\Q}\wedge d\,\Q \wedge e^{-ut\Q}. $$ \end{lem} {\bf
Proof:} Take $V_u= e^{-uQ} \, e^{-(1-u) (\Q+d\, \Q)}$. Then  $$\frac{d}{du }
V_u= -\Q\, e^{-u\Q} \, e^{-(1-u) (\Q+d\, \Q)} + e^{-u \Q} \, (\Q+ d\, \Q) \,
e^{-(1-u) (\Q+d\, \Q)}.$$ Integrating this from $0$ to $1$ yields $$ d e^{-\Q} :=
e^{-\Q} -e^{-(\Q+d\, \Q)} =  \int_0^1  du\, e^{-u \, \Q} d\Q\, e^{-(1-u)
(\Q+d\, \Q)}.$$ Replacing $\Q$ by $t\, \Q$ and making the adequate change of
variable then yields the result. \begin{prop} \label{prop:dchiQ}  For any
$A_0, \cdots, A_n \in \Omega(X, {\cal A})$, $$\left(d \, \tilde \chi_{n,
\Q}\right)(t)(A_0, \cdots, A_{n})= -t\,\sum_{j=1}^{n+1}\tilde \chi_{n+1,
\Q}\left( A_0, A_1,\cdots, A_{j-1},d\,   \Q,A_{j},  \cdots , A_{n}\right),$$
where $d \,  \Q$ stands at the $j$-th position.  \end{prop} {\bf Proof:} Since
any power  $e^{-u \Q}, 0<u<1$ lies in ${\cal I}$, so does the product\hfill\break\noindent $A_0\, e^{-u_0
\Q}\,  A_1\, e^{-u_1 \Q}\, \cdots \, A_{n} \, e^{-u_n \Q}$ lie in ${\cal I}$.\\
We can push
 $d$ through the trace $T$ in the subsequent computation and use 
Lemma \ref{lem:dchi} to express $d\, e^{-u\Q}$: \begin{eqnarray*} &{}& d\,
\left( T \left(A_0\, e^{-u_0\Q}\, A_1\, e^{-u_1\Q}\, \cdots A_{j} \,
\e^{-u_j\Q}\,  A_{j+1} \cdots  A_{n} \, e^{-u_n\Q}\right)\right)\\ &-&
\sum_{j=1}^n  T \left(A_0\, e^{-u_0\Q}\,  A_1\,e^{-u_1\Q}\, \cdots   d\, 
A_{j} \,   e^{-u_j\Q}\,  A_{j+1}\,\cdots \, \alpha_{n} \, e^{-u_n\Q}\right)\\
&= &T \left(d\, \left( A_0\,e^{-u_0\Q}\,  A_1\, e^{-u_1\Q} \cdots  A_{j} \,
e^{-u_j\Q}\,  A_{j+1} \cdots A_{n} \, e^{-u_n\Q}\right)\right)\\ &-& 
\sum_{j=1}^{n+1} T \left(A_0\, e^{-u_0\Q}\,  A_1\, e^{-u_1\Q} \cdots  d\, 
A_{j} \,   e^{-u_j\Q}\,  A_{j+1} \cdots A_{n} \, e^{-u_n\Q}\right)\\ &=&
\sum_{j=1}^{n+1} T\left( A_0\, e^{-u_0\Q}\, A_1\, e^{-u_1\Q}  \cdots  
A_{j-1}\, d\,\left(  e^{-u_{j-1}\Q}\right)\, A_j   \cdots A_{n} \, e^{-u_n\Q}
\right)\\ &=& -\sum_{j=1}^{n+1}\, u_{j-1} \, \,T\left( A_0\, e^{-u_0\Q}\,  
A_1\, e^{-u_1\Q} \cdots \right.\\ &{}&\left.\cdots  A_{j-1}\,  \left( \int_0^1
du\, e^{-u_{j-1} (1-u)\Q}\, d\, \Q\, e^{-u\, u_{j-1} \Q}\right)\, A_j\, 
e^{-u_j\Q}\, A_{j+1} \cdots A_{n} \, e^{-u_n\Q} \right).\\ \end{eqnarray*} 
Replacing $\Q$ by $t\, \Q$ we then integrate  over the simplex $\Delta_n$; 
the integration  $$\int_0^1 du\, e^{-u_{j-1} (1-u)\Q}\,d\,  \Q\,e^{-u_{j-1}
u\Q}$$ inside the above expression  gives rise to an integration on the simplex
$\Delta_{n+1}$ and yields the result.   \section{Comparing weighted trace cochains for different
weights} Let us  consider  a fixed pseudodifferential algebra $Cl(M, E)$,
where as before $\pi:E\to M$ is a hermitian vector bundle on a fixed closed
Riemannian manifold $M$, and a smooth family $\Q: x\mapsto Q_x\in Cl(M, E)$ of
 pseudodifferential operators with positive scalar leading symbol and constant
order $q$ parametrized by a smooth manifold $X$. Their leading symbol being
invertible, these operators are elliptic.   This gives rise to  a smooth
family of weighted pseudodifferential algebras  $$\left({\cal A}= Cl(M, E),
\Q\right):=\left({\cal A}, Q_x, x\in X\right).$$  Given  any $A_0, \cdots, A_n
\in C^\infty (X, {\cal A})$,  for any $ \e>0$, $ \tilde \chi_{ n,\Q}(\e)(A_0,
\cdots, A_n)$ is a smooth  function on $X$  defined at point $x\in X$ by  $
\tilde \chi_{ n,Q_x}(\e)(A_0, \cdots, A_n)$ and   $\bar \chi_{n, \Q}(A_0,
\cdots, A_n)$ is a smooth function on $X$  defined at point $x$ by $\bar
\chi_{n, Q_x}(A_0, \cdots, A_n)$. \\ Let us assume that $\Q$ is invertible.
Otherwise, provided ${\rm Ker} \,\Q$ defines a vector bundle, we can replace
$\Q$ by $\Q+ \pi_{Ker \, \Q}$ where $\pi_{Ker \, \Q}$ is otrhogonal projection
onto the kernel bundle.  We can therefore apply the result of the previous
section to every $Q_x, x\in X$, which yield that for  any $A_0, \cdots, A_n\in
C^\infty\left(X,Cl(\M,\E)\right)$, the map $z\mapsto \bar \chi_{n, \Q}(z)(A_0,
\cdots, A_{n})$ is meromorphic with simple poles. Its complex residue at
$z=0$, which is independent of $x\in X$  is given by: $$ {\rm Res}_{z=0}  \bar
\chi_{n, \Q}(z)(A_0, \cdots, A_{n})=\frac{1}{q}\, {\rm res} \left(A_0 \,A_1, 
\cdots ,A_n \right)$$ and its finite part, which we call the {\bf
$\Q$-weighted trace cochain } is a smooth function on $X$ given at each point
$x\in X$  by  \begin{eqnarray*} &{}& \chi_n^{Q_x}(A_0, \cdots, A_{n}) \\
&=&{\rm tr}^{Q_x}\left(A_0\,A_1 \cdots \,A_n \right)+\frac{1}{q} \sum_{\vert
k\vert=1}^{[\vert a\vert]+{\rm dim} M} \frac{(-1)^{\vert k\vert}(\vert
k\vert-1)!}{ (k+1)!} \, {\rm res} \left(A_0 \, A_1^{(k_1)}\,  \cdots  \,
A_n^{(k_n)}\, Q_x^{-\vert k\vert}\right),\\ \end{eqnarray*} where $\vert
a\vert= \sum_{i=0}^n {\rm ord}( A_i)$ and where for any multiindex
$k=(k_1, \cdots, k_n)$ we have set $\vert k\vert = k_1+\cdots +k_n$, $k!=
k_1!\cdots k_n!$. \begin{thm}\label{thm:dbarchi} The map  $  d \, \bar
\chi_{n, \Q}$  is holomorphic at zero and for any $A_0, \cdots, A_n \in
C^\infty(X, {\cal A})$, it is by: \begin{eqnarray*}
&{}& \left(d \,\chi_n^\Q\right)   (A_0, \cdots, A_{n} )=\lim_{z\to 0}\left(d
\,\bar \chi_{n, \Q}\right)(z) (\alpha_0, \cdots, \alpha_{n})\\ &=& 
\frac{1}{q}\cdot  \sum_{\vert k\vert=0}^{[\vert a\vert] +{\rm dim }M} 
\frac{(-1)^{\vert k\vert+1} k!}{ (k+1)!} \sum_{j=1}^{n+1 } {\rm res} \left(
A_0 \, A_1^{(k_1)}\,\cdots  A_{j-1}^{(k_{j-1})} \,   \left(d\,
\Q\right)^{(k_j)}\,A_j^{(k_{j+1})} \right.\\
&{}&\left.   \cdots\,  A_n^{(k_{n+1})} \Q^{-\vert k\vert-1}\right)\\
\end{eqnarray*} where we have set $\vert a\vert= \sum_{i=0}^n {\rm ord}
(A_i)$,  $[\vert a\vert]$ to be its integer part, and  $\vert k\vert =
k_1+\cdots +k_{n+1}$. \end{thm} {\bf Proof:} By Proposition \ref{prop:dchiQ}
 we have: \begin{eqnarray*}&{}& \left(d \,\tilde 
\chi_{n, \Q}(t)\right)(A_0, \cdots, A_{n}) \\
&=&-t \, \sum_{j=1}^{n+1} 
\tilde \chi_{n+1, \Q}(t) \left( A_0, A_1,\cdots,A_{j-1},d\, \Q, A_{j},  
\cdots , A_{n}\right), \end{eqnarray*} where we have inserted $d\, \Q$
at the $j$-th position.  It follows that  \begin{eqnarray*} &{}& \left(d \,
\bar \chi_{n, \Q}(z)\right)(A_0, \cdots, A_{n})\\ &=&- \frac{1}{\Gamma(z)}
\sum_{j=1}^{n+1} \int_0^\infty t^z  dt\, \tilde \chi_{ n+1,\Q}(t)(A_0,\cdots,
A_{j-1}, d\,    \Q,A _j\cdots, A_n)\, dt\\ &= & -\frac{1}{\Gamma(z)}  
\sum_{j=1}^{n+1}    \sum_{ k_1=0}^{\infty}\cdots \sum_{k_{n+1}=0}^\infty
\frac{(-1)^{\vert k\vert} }{ (k+1)!}   \int_0^\infty dt\,t^{z+\vert k\vert}\,
{\rm tr} \left( A_0 \, A_1^{(k_1)} \cdots\right.\\ &{}& \left.
A_{j-1}^{(k_{j-1})} \, \left( d\,  \Q\right)^{(k_j)}\,  A_{j}^{(k_{j+1})}
\cdots    A_{n}^{(k_{n+1})}\, e^{-t\Q}\right)\, dt\\ &= &  \sum_{j=1}^{n+1} 
\sum_{k_1=0}^{\infty}\cdots \sum_{k_{n+1}=0}^{\infty} \frac{(-1)^{\vert
k\vert+1} }{ (k+1)!}  \frac{\Gamma( z+\vert k\vert+1)}{\Gamma(z)}  {\rm TR}
\left( A_0 \,A_1^{(k_1)} \, \cdots \right.\\ &{}& \left. \cdots
A_{j-1}^{k_{j-1}}\, \left(d\, \Q\right)^{(k_j)}\, A_{j}^{(k_{j+1})}
\cdots    A_n^{(k_{n+1})}\, \Q^{ - ( z+\vert k\vert+1)} \right),\\
\end{eqnarray*}
where we have used Appendix B in the first equation. \\
 For any  multiindex $ k$ (including $\vert k\vert=0$),  $
\frac{\Gamma( z+\vert k\vert+1)}{\Gamma(z)}= \vert k\vert!\cdot z$. Since $$
{\rm TR} \left( A_0 \, A_1^{(k_1)}  \cdots A_{j}^{(k_{j-1})}\, \left(d\, \Q\right)^{(k_j)}\,
A_{j}^{(k_{j+1})}  \cdots   A_n^{(k_{n+1})} \Q^{ - ( z+\vert k\vert+1)}
\right)$$ has a simple pole at zero given by $$\frac{1}{q}  \, {\rm res}
\left( A_0 \, A_1^{(k_1)}  \cdots    A_{j}^{(k_{j-1})} \,
\left(d\Q\right)^{(k_j)}\,A_{j}^{(k_{j+1})} \cdots   A_n^{(k_{n+1})} \Q^{ -( \vert k\vert+1)}
\right),$$ it follows that $\left(d\, \bar \chi_{n, \Q}\right)(z)(A_0, \cdots,
A_{n})$  converges to  \begin{eqnarray*} &{}& \lim_{z\to 0}\left(d \, \bar
\chi_{n, \Q}(z)\right)(A_0, \cdots, A_{n}) \\ & =& \frac{1}{q}\cdot
\sum_{k_1=0}^{\infty}\cdots  \sum_{k_{n+1}=0}^{\infty}(-1)^{\vert k\vert+1}
\,\frac{k!}{(k+1)!} \sum_{j=1}^{n+1 }  {\rm res} \left( A_0 \,
A_1^{(k_1)}\cdots \right.\\ &{}& \left.\cdots  A_{j-1}^{(k_{j-1})} \,  
\left(d\Q\right)^{(k_j)}\,A_{j}^{(k_{j+1})}  \cdots A_n^{(k_{n+1})}
\Q^{-\vert k\vert-1}\right).\\ \end{eqnarray*} Let us  now check that the sum
over $k$ is finite. Since $d\, \Q$  has order bounded by $q$, the operator
valued $1$-form  $A_0 \,  A_1^{(k_1)}\, \cdots \,  A_{j-1}^{(k_{j-1})} \,  
\left(d\,\Q\right)^{(k_j)} \,A_{j}^{(k_{j+1})} \cdots\, 
A_n^{(k_{n+1})}
 \Q^{-\vert k\vert-1}$ has order with upper bound $\vert
a\vert+q +\vert k\vert (q-1) -q (\vert k \vert+1)= \vert a\vert-\vert k\vert$.
The sum over the multiindices $k$ therefore goes up to  $[\vert a\vert]+{\rm
dim} M$.  \begin{cor} Let $\left(Cl(M, E), Q_0\right)$ and $\left( Cl(M, E), 
Q_1\right)$ be two weighted pseudodifferential algebras with $Q_0$ and $Q_1$
of same order $q>0$.   Then  \begin{eqnarray*} &{}& \chi_{n}^{ Q_1}(A_0,
\cdots, A_n)- \chi_{n}^{ Q_0}(A_0, \cdots, A_n)\\ &=& \frac{1}{q}\cdot 
\sum_{\vert k\vert=0}^{[\vert a\vert] +{\rm dim }\,M} (-1)^{\vert k\vert+1}\, 
\frac{ k!}{ (k+1)!}\sum_{j=1}^{n+1 }   \int_0^1 dt\, {\rm res} \left( A_0 \,
A_1^{(k_1)}\,\cdots\right.\\ &{}& \left. \,   A_{j-1}^{(k_{j-1})} \,  
\left(\frac{d}{dt} Q_t\right)^{(k_j)}  \cdots\,  A_n^{(k_{n+1})} Q_t^{-\vert
k\vert-1}\right)\\ \end{eqnarray*} \end{cor} {\bf Proof:} This follows from
applying the above theorem to $X=[0, 1]$ and  a   smooth family $Q_t$
interpolating $Q_0$ and $Q_1$.  \begin{rk} The one parameter family   $Q_t:= Q_0^{1-t} Q_1^t,
t\in [0, 1]$ interpolates $Q_0$ and $Q_1$ and each $Q_t$ is an  elliptic
operator with positive scalar symbol $\sigma_L(Q_t)= \sigma_L(Q_0)^{1-t}\,
\sigma_L( Q_1)^t$. Since $$\frac{d}{dt} Q_t= -\log Q_0 \, Q_t +Q_0^{1-t}\,
\log Q_1 Q_1^t= -\log Q_0 \, Q_t +Q_t\, \log Q_1 ,$$   integrating on $[0, 1]$
we get: \begin{eqnarray*} &{}& \chi_{n, Q_1}(A_0, \cdots, A_n)- \chi_{n,
Q_0}(A_0, \cdots, A_n)\\ &=& \frac{1}{q}\cdot  \sum_{\vert k\vert=0}^{[\vert
a\vert] +{\rm dim }\,M} (-1)^{\vert k\vert+1}\,  \frac{ k!}{
(k+1)!}\sum_{j=1}^{n+1 }   \int_0^1 dt\, {\rm res} \left( A_0 \,
A_1^{(k_1)}\,\cdots\right.\\ &{}& \left. \,   A_{j-1}^{(k_{j-1})} \,   \left(
-\log Q_0 \, Q_t +Q_t\, \log Q_1\right)^{(k_j)}\,  A_j^{(k_{j+1})}  \cdots\,  A_n^{(k_{n+1})}
Q_t^{-\vert k\vert-1}\right)\\ &=& \frac{1}{q}\cdot  \sum_{\vert
k\vert=0}^{[\vert a\vert] +{\rm dim }\,M} (-1)^{\vert k\vert+1}\,  \frac{ k!}{
(k+1)!}\sum_{j=1}^{n+1 }   \int_0^1 dt\, {\rm res} \left( A_0 \,
A_1^{(k_1)}\,\cdots\right.\\ &{}& \left. \,   A_{j-1}^{(k_{j-1})} \, \left( -
\left( \log Q_0\right)^{(k_j)} \, Q_t +Q_t\, \left(\log Q_1
\right)^{(k_j)}\right)\,A_j^{(k_{j+1})} \cdots\,  A_n^{(k_{n+1})} Q_t^{-\vert
k\vert-1}\right)\\ &=& \frac{1}{q}\cdot  \sum_{\vert k\vert=0}^{[\vert a\vert]
+{\rm dim }\,M} (-1)^{\vert k\vert}\,  \frac{ k!}{ (k+1)!} \int_0^1 dt\, {\rm
res} \left( A_0 \, A_1^{(k_1)}\,\cdots\right.\\ &{}& \left. \,  
A_{j-1}^{(k_{j-1})} \,  \left( \log Q_0\right)^{(k_j)}A_j^{(k_{j+1}+1)}
\cdots\,  A_n^{(k_{n+1})} Q_t^{-\vert k\vert-1}\right)\\ &+&  \frac{1}{q}\cdot
 \sum_{\vert k\vert=0}^{[\vert a\vert] +{\rm dim }\,M} (-1)^{\vert k\vert}\, 
\frac{ k!}{ (k+1)!}  \int_0^1 dt\, {\rm res} \left( A_0 \,
A_1^{(k_1)}\,\cdots\right.\\ &{}& \left. \,   A_{j-1}^{(k_{j-1})} \,  \left(
\log Q_0\right)^{(k_j)}A_j^{(k_{j+1})} \, A_{j+1}^{(k_{j+2}+1)}\,
A_{j+2}^{(k_{j+3})}\, \cdots\,  A_n^{(k_{n+1})} Q_t^{-\vert k\vert-1}\right)\\
&+&  \frac{1}{q}\cdot  \sum_{\vert k\vert=0}^{[\vert a\vert] +{\rm dim }\,M}
(-1)^{\vert k\vert}\,  \frac{ k!}{ (k+1)!}  \int_0^1 dt\, {\rm res} \left( A_0
\, A_1^{(k_1)}\,\cdots\right.\\ &{}& \left. \,   A_{j-1}^{(k_{j-1})} \, 
\left( \log Q_0\right)^{(k_j)}A_j^{(k_{j+1})} \, A_{j+1}^{(k_{j+2})} \cdots\, 
A_n^{(k_{n+1})} Q_t^{-\vert k\vert}\right)\\ &+&\frac{1}{q}\cdot  \sum_{\vert
k\vert=0}^{[\vert a\vert] +{\rm dim }\,M} (-1)^{\vert k\vert}\,  \frac{ k!}{
(k+1)!} \int_0^1 dt\, {\rm res} \left( A_0 \, A_1^{(k_1)}\,\cdots\right.\\
&{}& \left. \,   A_{j-1}^{(k_{j-1}+1)} \,  \left( \log
Q_1\right)^{(k_j)}A_j^{(k_{j+1})} \cdots\,  A_n^{(k_{n+1})} Q_t^{-\vert
k\vert-1}\right)\\ &+&  \frac{1}{q}\cdot  \sum_{\vert k\vert=0}^{[\vert
a\vert] +{\rm dim }\,M} (-1)^{\vert k\vert}\,  \frac{ k!}{ (k+1)!}  \int_0^1
dt\, {\rm res} \left( A_0 \, A_1^{(k_1)}\,\cdots A_{j-2}^{(k_{j-2}+1)}\right.\\ &{}& \left. \,  
A_{j-1}^{(k_{j-1})} \,  \left( \log Q_1\right)^{(k_j)}A_j^{(k_{j+1})} \,
A_{j+1}^{(k_{j+2})}\, A_{j+2}^{(k_{j+3})}\, \cdots\,  A_n^{(k_{n+1})}
Q_t^{-\vert k\vert-1}\right)\\&+&\cdots+\\ &+&  \frac{1}{q}\cdot  \sum_{\vert
k\vert=0}^{[\vert a\vert] +{\rm dim }\,M} (-1)^{\vert k\vert}\,  \frac{ k!}{
(k+1)!} \int_0^1 dt\, {\rm res} \left( A_0 \, A_1^{(k_1)}\,\cdots\right.\\
&{}& \left. \,   A_{j-1}^{(k_{j-1})} \,  \left( \log
Q_1\right)^{(k_j)}A_j^{(k_{j+1})} \, A_{j+1}^{(k_{j+2})} \cdots\, 
A_n^{(k_{n+1})} Q_t^{-\vert k\vert}\right)\\ \end{eqnarray*} 
where we have used the fact that $[Q_t, \log \, Q_0 \, Q_t]= [Q_t, \log Q_0] \, Q_t$ and 
$ [ Q_t,Q_t\, \log Q_1]= Q_t \,[Q_t, \log Q_1]$.  Since $Q_t$ has
scalar leading symbol, and since $\sigma(\log Q_t)= q\, log \vert \xi\vert +
\sigma_0(\log Q_t)$ where $\sigma_0(\log Q_t)$ is a classical (local) symbol
of  of order $0$,   $ \left( \log Q_0\right)^{(k_j)}$ has order  $k_j(q-1)$.
The order of the expression inside the Wodzicki residue is therefore $\vert
a\vert +\vert k\vert \,(q-1)- q\, \vert k\vert= \vert a\vert -\vert k\vert$. As a consequence,  the sum over  $\vert k\vert$ stops at $[\vert
a\vert]+{\rm dim}\, M$ since beyond that   $\vert a\vert -\vert k\vert<-{\rm
dim}\, M$.  \end{rk}

\section{Fibrations of weighted trace algebras }
Let  now ${\cal A}\to X$  be a smooth fibration  of  associative algebras over $\C$  based on  a manifold $X$ equipped with a connection $\nabla$. The previous section took care of the case when the fibration is trivial and $\nabla= d$.  Let  $C^n\left( {\cal  A}\right)\to X$
denote the corresponding fibration  of  spaces of $n$-cochains on ${\cal
A}$. It is  equipped with the induced connection:
\begin{eqnarray*}
[\nabla\, \chi_n](\alpha_0, \cdots, \alpha_n)&:=& d\left[ \chi_n(\alpha_0, \cdots, \alpha_n)\right]\\
&-&\sum_{j=0}^{n} (-1)^{\vert \alpha_0\vert +\cdots +\vert \alpha_{j-1}\vert } \chi_n
(\alpha_0, \cdots, \nabla\, \alpha_j, \alpha_{j+1}, \cdots, \alpha_n),
\end{eqnarray*}
 where $\alpha_0, \cdots, \alpha_n \in \Omega(X,{\cal A})$, 
the space of forms on $X$ with values in ${\cal A}$. It takes $k$-forms with values in $n$-cochains to $k+1$-forms with values in $n$-cochains.
\begin{defn} 
 A smooth fibration of  weighted tracial algebras is a triple  $\left({\cal A}, \Q, T\right)$ where $T:{\cal I}\to X\times \C$ is a bundle linear mophism acting on a bundle ${\cal I}\to X$ of fibrewise ideals of ${\cal A}$ and $\Q$ an element of $\Omega^{2p}(X, {\cal A} )$ for some non  negative integer $p$, that satisfies the following requirements:
\begin{enumerate}
\item $T$ is fibrewise a  trace on ${\cal I}$,
\item  For each $x\in X$, and for any $U_1, \cdots, U_{2p}\in T_xX$,  $\left({\cal A}_x, Q_x(U_1, \cdots, U_{2p}), T_x\right)$ is a weighted trace algebra,
\end{enumerate}  
where we have extended $T$ to ${\cal I}$-valued forms on $X$. $\Q$ is called the weight.
 \\
\begin{rk}
When the fibration ${\cal A}\to X$ is trivial, $T$ can be taken constant so 
that choosing $p=0$,  we recover the framework of the previous section, namely $\left({\cal A}, \Q, T\right)$ where ${\cal A}$ is a fixed topological unital associative  algebra, $T$ a trace on a non trivial ideal ${\cal I}$ of ${\cal A}$ and $\Q: X\to C^\infty(X, {\cal A})$ a smooth family of weights. 
\end{rk}
To a smooth fibration of weighted tracial algebras we  associate smooth families of cochains:
\begin{eqnarray*}
&{}&\chi_{ n,\Q}(\alpha_0, \cdots,
\alpha_n)\\
&:=&\int_{\Delta_n}du_0\cdots du_n \, T\left( \alpha_0\wedge 
e^{-u_0\Q}\wedge  \alpha_1\wedge  e^{-u_1\Q} \wedge \cdots \wedge
\alpha_{n-1}\wedge    e^{-u_{n-1}\Q}\wedge \alpha_n\wedge e^{-u_n\Q}
\right)
\end{eqnarray*}
which are  fibrewise defined using $T_x$ on the fibre. \end{defn} 
\begin{lem}\label{lem:nabla} If the connection $\nabla$ is induced by a
principal bundle connection so that locally, $\nabla= d +{\rm ad}_\theta$
where $\theta$ is a local ${\cal A}$-valued one form, then  $$\nabla\, 
e^{-u\Q}= -u \int_0^1 dt \,e^{-u(1-t)\Q}\wedge  \nabla\,   \Q \wedge e^{-ut\Q}.
$$ \end{lem} {\bf Proof:}  Combining  Lemma \ref{lem:dchi}  with   Lemma
\ref{lem:ad} yields: \begin{eqnarray*}
\nabla\,  e^{-u\Q}&=& d \,  e^{-u\Q}+ [\theta, e^{-u\Q}]\\
 &=& -u \,\int_0^1 dt\,e^{-u(1-t)\Q}\wedge  d   \Q \wedge e^{-ut\Q}
 -u \int_0^1 dt\,e^{-u(1-t)\Q}\wedge  {\rm ad}_\theta \Q \wedge e^{-ut\Q}\\
&=&-u \,\int_0^1 dt\,e^{-u(1-t)\Q}\wedge  d  \Q \wedge e^{-ut\Q}
 -u \int_0^1 dt \,e^{-u(1-t)\Q}\wedge  [\theta,  \Q] \wedge e^{-ut\Q}\\
&=&-u \,\int_0^1 dt \,e^{-u(1-t)\Q}\wedge  \nabla  \Q \wedge e^{-ut\Q}. 
\end{eqnarray*}
As a consequence, 
\begin{prop}\label{prop:nabla}
Given a smooth fibration of weighted trace algebras $({\cal A}, \Q,T)$  
based on $X$ equipped with a connection $\nabla$ such that $[\nabla\, T]$
vanishes on ${\cal I}$, then  for any $\alpha_0, \cdots, \alpha_n \in
\Omega(X, {\cal A})$, \begin{eqnarray*} &{}& \left[\nabla \, \chi_{n,
\Q}\right](\alpha_0, \cdots, \alpha_{n})\\ &=&
\sum_{j=1}^{n+1} 
(-1)^{\vert \alpha_0\vert +\cdots +\vert \alpha_{j-1}\vert \, +1}\,\chi_{n+1, \Q}\left( \alpha_0, \alpha_1,\cdots, \alpha_{j-1},\nabla \, \Q,\alpha_{j},  \cdots , \alpha_{n}\right)
\end{eqnarray*}
where $\nabla \,  \Q$ stands at the $j$-th position. 
\end{prop}
{\bf Proof:}
Since any $e^{-u\Q}, 0<u<1$ lies in ${\cal I}$, so does $\alpha_0\wedge e^{-u_0\Q}\wedge  \alpha_1\wedge e^{-u_1\Q}\wedge \cdots  \wedge \alpha_{j} \wedge   e^{-u_j\, \Q}\wedge  \alpha_{j+1}\wedge \cdots
\wedge \alpha_{n} \wedge e^{-u_n\Q}$ lie in ${\cal I}$. Since $\nabla T$ vanishes on ${\cal I}$, it follows that we can push $\nabla$ through the trace $T$ in the subsequent computation using Lemma \ref{lem:nabla} to express $\nabla 
e^{-u_j\Q}$:
\begin{eqnarray*}
&{}&\nabla\left( T \left(\alpha_0\cdot e^{-u_0\Q}\wedge 
 \alpha_1\cdot e^{-u_1\Q}\wedge \cdots  \wedge \alpha_{j} \wedge  
e^{-u_j\Q}\wedge  \alpha_{j+1}\wedge \cdots \wedge \alpha_{n} \cdot
e^{-u_n\Q}\right)\right)\\
 &-&  \sum_{j=1}^{n+1}(-1)^{\vert \alpha_0\vert 
+\cdots +\vert \alpha_{j-1}\vert} T \left(\alpha_0\cdot e^{-u_0\Q}\wedge 
\alpha_1\wedge e^{-u_1\Q}\wedge \cdots  \wedge \nabla \alpha_{j} \wedge  
e^{-u_j\Q}\wedge  \alpha_{j+1}\wedge\cdots \right.\\
&{}& \left.\cdots \wedge \alpha_{n} \wedge
e^{-u_n\Q}\right)\\ &=& T \left(\nabla\left( \alpha_0\wedge e^{-u_0\Q}\wedge 
\alpha_1\wedge e^{-u_1\Q}\wedge \cdots  \wedge \alpha_{j} \wedge  
e^{-u_j\Q}\wedge  \alpha_{j+1}\wedge\right.\right.\\
&{}&\left.\left. \cdots \wedge \alpha_{n} \wedge
e^{-u_n\Q}\right)\right)\\
 &-&  \sum_{j=1}^{n+1}(-1)^{\vert \alpha_0\vert 
+\cdots +\vert \alpha_{j-1}\vert} T \left(\alpha_0\wedge e^{-u_0\Q}\wedge 
\alpha_1\cdot e^{-u_1\Q}\wedge\cdots\right.\\
&{}&\left.  \cdots  \wedge \nabla \alpha_{j} \wedge  
e^{-u_j\Q}\wedge  \alpha_{j+1}\wedge \cdots \wedge \alpha_{n} \cdot
e^{-u_n\Q}\right)\\ &=& \sum_{j=1}^{n+1}(-1)^{\vert \alpha_0\vert  +\cdots
+\vert \alpha_{j-1} \vert} T\left( \alpha_0\wedge e^{-u_0\Q}\wedge 
\alpha_1\cdot e^{-u_1\Q} \wedge \cdots\right.\\ &{}&\left. \cdots   \wedge
\alpha_{j-1}\wedge \nabla e^{-u_{j-1}\Q}\wedge\alpha_j\wedge   \cdots \wedge
\alpha_{n} \wedge e^{-u_n\Q} \right)\\ &=& \sum_{j=1}^{n+1}(-1)^{\vert
\alpha_0\vert  +\cdots +\vert \alpha_{j-1}\vert+1} u_{j-1} \, \,T\left(
\alpha_0\wedge e^{-u_0\Q}\wedge   \alpha_1\wedge e^{-u_1\Q}\wedge \cdots \wedge
\alpha_{j-1}\right.\\ &{}&\left. \wedge \left( \int_0^1 dt\,e^{-u_{j-1}
(1-t)\Q}\wedge \nabla \Q\wedge e^{-u_{j-1} t\Q}\right)\wedge \alpha_j\wedge
e^{-u_j\Q}\wedge  \alpha_{j+1}\wedge\right.\\
&{}& \left.\cdots \wedge \alpha_{n} \wedge
e^{-u_n\Q} \right).\\ \end{eqnarray*}  Integrating over the simplex
$\Delta_n$, the integration  $\int_0^1 dt\,  e^{-u_{j-1} (1-t)\Q}\wedge \nabla
\Q\wedge e^{-u_{j-1} t\Q}$ inside the above expression  gives rise to an
integration on the simplex $\Delta_{n+1}$ and yields the result. 
\begin{cor}\label{cor:nabla} Given a smooth fibration of weighted trace
algebras $({\cal A}, \Q,T)$  based on $X$ equipped with a connection $\nabla$
such that $[\nabla\, T]$ vanishes on ${\cal I}$, then  for any $\alpha_0,
\cdots, \alpha_n \in \Omega(X, {\cal A})$, \begin{eqnarray*} &{}&  d
\left(\chi_{n, \Q}(\alpha_0, \cdots, \alpha_{n})\right)\\ &=& \sum_{j=1}^{n} 
(-1)^{\vert \alpha_0\vert +\cdots \vert \alpha_{j-1}\vert}\,  \chi_{n,
\Q}\left( \alpha_0, \alpha_1,\cdots, \alpha_{j-1},\nabla
\alpha_{j},\alpha_{j+1},  \cdots , \alpha_{n}\right)\\ &+&\sum_{j=1}^{n+1} 
(-1)^{\vert \alpha_0\vert +\cdots +\vert \alpha_{j}\vert \, +1}\,  \chi_{n+1,
\Q}\left( \alpha_0, \alpha_1,\cdots, \alpha_{j-1},\nabla  \Q,\alpha_{j}, 
\cdots , \alpha_{n}\right)\\ \end{eqnarray*} where $\nabla \, \Q$ stands at
the $j$-th position \end{cor} {\bf Proof:} This follows from the above
proposition combined with the fact that: \begin{eqnarray*} &{}&  d\left(\chi_{n,
\Q}(\alpha_0, \cdots, \alpha_{n})\right)\\ &=& \sum_{j=1}^{n} 
(-1)^{\vert \alpha_0\vert +\cdots +\vert \alpha_{j-1}\vert}\,
 \chi_{n, \Q}\left( \alpha_0, \alpha_1,\cdots, \alpha_{j-1},\nabla
 \alpha_{j},  \cdots , \alpha_{2k+1}\right)\\
&+ & \left[\nabla \, \chi_{n,
\Q}\right](\alpha_0, \cdots, \alpha_{n}). \end{eqnarray*}
\section{Weighted trace cochains for families of pseudodifferential operators}
Consider a smooth fibration of smooth closed Riemannian manifolds 
$\pi:\M\to X$ with fibre $M_x$ above $x\in X$  and a smooth (possibly
$\Z_2$-graded) vector bundle $\E$ on  $\M$. Let ${\cal E}:=\pi_*\E$
denote the  (possibly $\Z_2$-graded) infinite rank vector bundle with
fibre above $x$ given by $C^\infty(M_x, \E_{\vert_{M_x}})$. Assuming that $\E$
is a hermitian bundle and combining the hermitian structure on
$\E_{\vert_{M_x}}$ with the Riemannian structure on the fibres $M_x$ yields an
$L^2$ structure on the fibres of ${\cal E}$.\\  This geometric setting also
gives rise to  a  smooth fibration   ${\cal A}:= Cl(\M, \E)\to X$  of vertical
classical pseudodifferential operators with fibre above $x\in X$ given by
${\cal A}_x:= Cl\left(M_x, \E_{\vert_{M_x}}\right)$.  There is a natural 
fibration of smoothing operators  ${\cal I}\to X$ with fibre above $x\in X$
given by the algebra ${\cal I}_x$ of smoothing operators on $M_x$ and  there
is a smooth fibre bundle morphism ${\rm tr}: {\cal I} \to X\times \C $ defined
by the ordinary trace ${\rm tr}_x$ on the ideal ${\cal I}_x$. \\  Let $\Q\in
\Omega^{2p}(X, {\cal A})$ for some non negative  integer $p$, be a $Cl(\M,
\E)$-valued even form such that  $$\forall x\in X, \, \forall \,U_1, \cdots, U_{2p}\in
T_xX, \quad \Q(U_1, \cdots, U_{2p})\quad {\rm is} \quad {\rm a} \quad {\rm
weight} \quad {\rm of} \quad  {\rm constant}\quad {\rm order} \, q. $$ 
$\left({\cal A}, \Q, {\rm tr}\right)$ defines a smooth fibration of weighted
trace algebras since  \begin{enumerate} \item  ${\rm tr}$ is fibrewise a 
trace on ${\cal I}$, \item  for each $x\in X$, and for any $U_1, \cdots,
U_{2p}\in T_xX$,  $\left({\cal A}_x, Q_x(U_1, \cdots, U_{2p}), {\rm
tr}_x\right)$ is a weighted trace algebra.  \end{enumerate}  We get this way a
smooth fibration  $\left({\cal A}, \Q\right)$ of weighted pseudodifferential
algebras  $\left({\cal A}_x, Q_x \right), \, x\in X$.\\  \begin{defn} Given 
any $\alpha_0, \cdots, \alpha_n \in \Omega(X, {\cal A})$ set for any $ \e>0$
\begin{eqnarray*} \tilde \chi_{ n,\Q}(\e)(\alpha_0, \cdots,
\alpha_n)&:=&\int_{\Delta_n}du_0\cdots du_n  \, {\rm tr}\left(  \alpha_0\wedge
  e^{-\e u_0\, \Q} \wedge \alpha_1\wedge  e^{ -\e u_1\, \Q} \cdots\right.\\
&{}&\left. \alpha_{n-1} \wedge  e^{-\e u_{n-1}\, \Q}\wedge \alpha_n\wedge 
e^{-\e u_{n}\, \Q}\right)\\ \end{eqnarray*} and let $\bar \chi_{n, Q}$ be its
Mellin transform: \begin{eqnarray*}  &{}&\bar \chi_{ n,\Q}(z)(\alpha_0, \cdots,
\alpha_n) \\
&:=&\frac{1}{\Gamma(z)}\, \int_0^\infty dt\,  t^{z-1} \,
\int_{\Delta_n}\, du_0\cdots du_n  \, {\rm tr}\left(  \alpha_0\wedge   e^{-t 
u_0\, \Q} \wedge \alpha_1\wedge  e^{ -t  u_1\, \Q} \cdots\right.\\ &{}&\left.
\alpha_{n-1} \wedge  e^{-t u_{n-1}\, \Q}\wedge \alpha_n\wedge  e^{-t u_{n}\,
\Q}\right).\\ \end{eqnarray*} \end{defn} A  connection $\nabla$ on ${\cal E}$
induces a connection  $\nabla^{\rm End}$ on $Cl(\M, \E)$ which relates to
$\nabla$ by  $\nabla^{{\rm End}} \alpha= \left[\nabla, \alpha\right], \,
\forall \alpha\in \Omega(X, Cl\left(\M, \E)\right)$ where the bracket is a
$\Z_2$-graded bracket.  \begin{prop}\label{prop:nablachi} For any $\alpha_0,
\cdots, \alpha_n \in  \Omega(X, Cl(\M, \E))$, \begin{eqnarray*} &{}&
\left(\nabla \,  \tilde \chi_{ n,\Q}\right)(t)(\alpha_0,\cdots, \alpha_n)\\
&=& t  \,\sum_{j=1}^{n+1} (-1)^{\vert \alpha_0\vert+\cdots +\vert
\alpha_{j-1}\vert+1} \tilde \chi_{ n+1,\Q}(\alpha_0,\cdots,
\alpha_{j-1},\nabla^{End}   \Q,\alpha_j\cdots, \alpha_n)\\ \end{eqnarray*}
where $\nabla^{End} \Q $ is at the $j$-th entry.  \end{prop} {\bf Proof:} 
Since ${\cal I}$ is equipped with the ordinary trace ${\rm tr}$ which 
commutes with exterior differentiation and vanishes on brackets, for all
$\alpha\in \Omega(X, {\cal I})$ we have \begin{eqnarray*} \left(\nabla\,{\tr}
\right) (\alpha)&:=&  d {\rm tr} (\alpha)- {\rm tr} \left(\nabla^{\rm End}
\alpha\right)\\ &=&  {\rm tr} (  d\, \alpha)- {\rm tr} \left([\nabla,
\alpha]\right)\\ &=& {\rm tr} (  d\, \alpha) + {\rm tr} (  [\Theta , \alpha])-
{\rm tr} \left([\nabla, \alpha]\right)\\ &=& 0\\ \end{eqnarray*} where we have
locally written $\nabla= d+\Theta$, with $\Theta$ a local one form with values
in $Cl(M, E)$. Hence $\nabla$ commutes with ${\rm tr} $ on ${\cal I}$. 
Applying 
Proposition \ref{prop:nabla}  to  $t\, \Q$
then yields the result. \\
When $\Q$ is a family of invertible operators we can apply fibrewise the results of the previous
sections to prove the following result. 
 \begin{thm}\label{thm:nablachiQ} Given
any $\alpha_0, \cdots, \alpha_n\in \Omega\left(X,Cl(\M,\E)\right)$, the map
$z\mapsto \bar \chi_{n, \Q}(z)(\alpha_0, \cdots, \alpha_{n})$ is meromorphic
with simple poles. Its complex residue at $z=0$ is given by: $$ {\rm
Res}_{z=0}  \left(\bar \chi_{n, \Q}(z)(\alpha_0, \cdots, \alpha_{n})\right)=\frac{1}{q}\,
{\rm res} \left(\alpha_0 \wedge \alpha_1 \wedge \cdots \wedge\alpha_n
\right)$$ and its finite part, which we call the {\bf $Q$-weighted trace
cochain }  by  \begin{eqnarray*}  \chi_n^\Q(\alpha_0, \cdots, \alpha_{n})
&:= &{\rm fp}_{z=0}  \bar \chi_{n, \Q}(z)(\alpha_0, \cdots, \alpha_{n})\\ &=
&{\rm tr}^{\Q}\left(\alpha_0\wedge  \alpha_1 \cdots \wedge
\alpha_n\right)\\
&+&\frac{1}{q} \sum_{\vert k\vert=1}^{[\vert a\vert]+{\rm dim}
M} \frac{(-1)^{\vert k\vert}(\vert k\vert-1)!}{ (k+1)!} \, {\rm res}
\left(\alpha_0 \wedge \alpha_1^{(k_1)}\wedge  \cdots \wedge
\alpha_n^{(k_n)}\wedge \Q^{-\vert k\vert}\right),\\ \end{eqnarray*} where
$\vert a\vert= \sum_{i=0}^n {\rm ord}( \alpha_i)$ and where for any multiindex
$k=(k_1, \cdots, k_n)$ we have set $\vert k\vert = k_1+\cdots +k_n$, $k!=
k_1!\cdots k_n!$. \\ Furthermore the map  $ \left( \nabla\, \bar \chi_{n,
\Q}(z)\right)$  is holomorphic at zero and for any $\alpha_0, \cdots, \alpha_n
$ in $\Omega(X, Cl(\M, \E))$ \begin{eqnarray*} &{}&  \left(\nabla
\,\chi_n^\Q\right)   (\alpha_0, \cdots, \alpha_{n} )\\ &=&\lim_{z\to
0}\left(\nabla \,\bar \chi_{n, \Q}(z)\right) (\alpha_0, \cdots, \alpha_{n})\\
&=&  \frac{1}{q}\cdot  \sum_{\vert k\vert=0}^{[\vert a\vert] +{\rm dim }M} 
\frac{ k!}{ (k+1)!}\sum_{j=1}^{n+1 } (-1)^{\vert \alpha_0\vert +\cdots + \vert
\alpha_{j-1}\vert+\vert  k\vert +1}  {\rm res} \left( \alpha_0 \wedge
\alpha_1^{(k_1)}\wedge \cdots\right.\\ &{}& \left. \cdots\wedge \,  
\alpha_{j-1}^{(k_{j-1})} \wedge   \left(\nabla^{End}\Q\right)^{(k_j)}\wedge\alpha_j^{({k_j+1})} \wedge  
\cdots\wedge  \alpha_n^{(k_{n+1})} \Q^{-\vert k\vert-1}\right)\\
\end{eqnarray*} where we have set $\vert a\vert= \sum_{i=0}^n {\rm ord}
(\alpha_i)$, and $[\vert a\vert]$ to be  its integer part, and  $\vert k\vert =
k_1+\cdots +k_{n+1}$. \end{thm} {\bf Proof:} The first part of the theorem
follows from Proposition \ref{prop:barchi} applied to each fibre.  Proposition \ref{prop:nablachi} yields the second part of the theorem.   By
Propostion \ref{prop:nablachi} we have: \begin{eqnarray*} 
&{}&\left(\nabla\,\tilde  \chi_{n, \Q}(t)\right)(\alpha_0, \cdots,
\alpha_{n})\\
&=&t \, \sum_{j=1}^{n+1} (-1)^{\vert \alpha_0\vert +\cdots + \vert
\alpha_{j-1}\vert+ 1} \tilde \chi_{n+1, \Q}(t) \left( \alpha_0,
\alpha_1,\cdots,\alpha_{j-1},\nabla^{End} \Q, \alpha_{j},   \cdots ,
\alpha_{n}\right),\nonumber \end{eqnarray*} where we have inserted
$\nabla^{End} \Q$ at the $j$-th position.  It follows that \begin{eqnarray*}
&{}& \left(\nabla \, \bar \chi_{n, \Q}(z)\right)(\alpha_0, \cdots,
\alpha_{n})\\ &=& \frac{1}{\Gamma(z)} \sum_{j=1}^{n+1} (-1)^{\vert
\alpha_0\vert +\cdots + \vert \alpha_{j-1}\vert+1}\int_0^\infty t^z  dt\,
\tilde \chi_{ n+1,\Q}(\alpha_0,\cdots, \alpha_{j-1},\nabla^{End}  
\Q,\alpha_j,\cdots, \alpha_n)\\ &= & \frac{1}{\Gamma(z)}   \sum_{j=1}^{n+1} 
(-1)^{\vert \alpha_0\vert +\cdots + \vert \alpha_{j-1}\vert+1}  \sum_{
k_1=0}^{\infty}\cdots \sum_{k_{n+1}=0}^\infty \frac{(-1)^{\vert k\vert} }{
(k+1)!}   \int_0^\infty dt\,t^{z+\vert k\vert}\, {\rm tr} \left( \alpha_0
\wedge\alpha_1^{(k_1)} \wedge \cdots\right.\\ &{}& \left.\cdots\wedge
\alpha_{j-1}^{(k_{j-1})} \wedge \left( \nabla^{End} \Q\right)^{(k_j)}\wedge 
\alpha_{j}^{(k_{j+1})} \wedge\cdots\wedge    \alpha_{n}^{(k_{n+1})}\wedge
e^{-t\Q}\right)\\ &= &   \sum_{j=1}^{n+1}  (-1)^{\vert \alpha_0\vert +\cdots +
\vert \alpha_{j-1}\vert+1}  \sum_{k_1=0}^{\infty}\cdots
\sum_{k_{n+1}=0}^{\infty} \frac{(-1)^{\vert k\vert} }{ (k+1)!}  \frac{\Gamma(
z+\vert k\vert+1)}{\Gamma(z)}  {\rm TR} \left( \alpha_0 \wedge
\alpha_1^{(k_1)} \wedge \cdots \right.\\ &{}& \left.
\cdots \wedge\alpha_{j-1}^{k_{j-1}}\wedge \left(\nabla^{End}
\Q\right)^{(k_j)}\wedge \alpha_{j}^{(k_{j+1})}\wedge \cdots \wedge  
\alpha_n^{(k_{n+1})}\wedge \Q^{ - ( z+\vert k\vert+1)} \right).\\
\end{eqnarray*} For any  multiindex $ k$ (including $\vert k\vert=0$),  $
\frac{\Gamma( z+\vert k\vert+1)}{\Gamma(z)}= \vert k\vert!\cdot z$. Since $$
{\rm TR} \left( \alpha_0 \wedge \alpha_1^{(k_1)}\wedge  \cdots \wedge 
\left(\nabla^{End}\Q\right)^{(k_j)}\wedge \alpha_{j}^{(k_{j+1})}\wedge  \cdots
\wedge  \alpha_n^{(k_{n+1})}\wedge \Q^{ - ( z+\vert k\vert+1)} \right)$$ has a
simple pole at zero given by $$\frac{1}{q}  \, {\rm res} \left( \alpha_0
\wedge \alpha_1^{(k_1)} \wedge \cdots \wedge   \alpha_{j}^{(k_{j-1})} \wedge
\left(\nabla^{End}\Q\right)^{(k_j)}\wedge \alpha_j^{(k_j+1)}\wedge \cdots  \wedge  \alpha_n^{(k_{n+1})}\wedge 
\Q^{ -( \vert k\vert+1)} \right),$$ it follows that $\left(\nabla\, \bar
\chi_{n, \Q}(z)\right)(\alpha_0, \cdots, \alpha_{n})$  converges to 
\begin{eqnarray*} &{}& \lim_{z\to 0}\left(\nabla \, \bar \chi_{n,
\Q}(z)\right)(\alpha_0, \cdots, \alpha_{n}) \\ & =& \frac{1}{q}\cdot
\sum_{k_1=0}^{\infty}\cdots  \sum_{k_{n+1}=0}^{\infty}(-1)^{\vert k\vert}
\,\frac{k!}{(k+1)!} \sum_{j=1}^{n+1 } (-1)^{\vert \alpha_0\vert +\cdots +
\vert \alpha_{j-1}\vert+1}  {\rm res} \left( \alpha_0 \wedge
\alpha_1^{(k_1)}\wedge\cdots \right.\\ &{}& \left.\cdots\wedge  
\alpha_{j-1}^{(k_{j-1})} \wedge   \left(\nabla^{End}\Q\right)^{(k_j)} \wedge \alpha_j^{(k_j+1)}\wedge
\cdots\wedge  \alpha_n^{(k_{n+1})}\wedge \Q^{-\vert k\vert-1}\right).\\
\end{eqnarray*} Let us  now check that the sum over $k$ is finite. Since
$\nabla^{End} \Q$  has order bounded by $q$, the operator valued form 
$\alpha_0 \wedge  \alpha_1^{(k_1)}\wedge \cdots \wedge  
\alpha_{j-1}^{(k_{j-1})} \wedge   \left(\nabla^{End}\Q\right)^{(k_j)} \wedge  \alpha_j^{(k_j+1)}\wedge 
\cdots\wedge  \alpha_n^{(k_{n+1})}\wedge \Q^{-\vert k\vert-1}$ has order with
upper bound $\vert a\vert+q +\vert k\vert (q-1) -q (\vert k \vert+1)= \vert
a\vert-\vert k\vert$. The sum over the multiindices $k$ therefore goes up to 
$[\vert a\vert]+{\rm dim} M$.  \begin{rk}  For $n=0$ this yields
\begin{eqnarray*} \left(\nabla {\rm tr}^\Q\right)(\alpha)&=& \left(\nabla
\chi_0^\Q\right)(\alpha)\\ &= &\sum_{k=0}^{[a]+{\rm dim} M} \frac{(-1)^{\vert
\alpha\vert +k+1}\, k!}{q\, (k+1)!} \, {\rm res} \left(\alpha \wedge 
\left(\nabla^{End} \Q\right)^{(k)}\wedge \Q^{-k-1}\right). \end{eqnarray*} It
compares with    a formula derived in  \cite{CDMP} by other means and used in
\cite{PR}: $$\left(\nabla \chi_0^\Q\right)(\alpha)= \frac{(-1)^{\vert
\alpha\vert +1}}{q} {\rm res} \left(\alpha \wedge  \nabla^{End} \log \Q\right)$$ which is more compact but more awkward to handle because of
the presence of a logarithm.  \end{rk} \begin{cor}\label{cor:dchinQalpha}
Given any $\alpha_0, \cdots, \alpha_n\in \Omega\left(X,Cl(\M,\E)\right)$,
\begin{eqnarray*} &{}&  d\,\left( \chi_n^\Q  (\alpha_0, \cdots, \alpha_{n}
)\right)\\ &=& \sum_{j=0}^n (-1)^{\vert \alpha_0\vert +\cdots+
\vert\alpha_{j-1}\vert} \chi_n^\Q  (\alpha_0, \cdots,\alpha_{j-1},
\nabla^{End}\alpha_j,\alpha_{j+1},  \cdots  \alpha_{n} )\\ &+&
\frac{1}{q}\cdot \sum_{\vert k\vert=0}^{[\vert a\vert] +{\rm dim}\, M} \,
\frac{k!}{(k+1)!} \sum_{j=1}^{{\rm dim}\, M+1 } (-1)^{\vert \alpha_0\vert
+\cdots + \vert \alpha_{j-1}\vert+\vert  k\vert +1}  {\rm res} \left( \alpha_0
\wedge \alpha_1^{(k_1)}\wedge \cdots\right.\\ &{}& \left.\cdots \wedge  \,
\alpha_{j-1}^{(k_{j-1})} \wedge   \left(\nabla^{End}\Q\right)^{(k_j)} 
\wedge \alpha_{j}^{(k_{j+1})}\wedge\cdots\wedge  \alpha_n^{(k_{n+1})} \wedge\Q^{-\vert k\vert-1}\right).\\
\end{eqnarray*} In particular,  when $[a]<-n$, then $\nabla$ "commutes" with
$\chi_n^\Q$: \begin{eqnarray*} &{}&  d\,\left( \chi_n^\Q  (\alpha_0, \cdots,
\alpha_{n} )\right)\\ &=& \sum_{j=1}^n (-1)^{\vert \alpha_0\vert +\cdots+
\vert\alpha_{j-1}\vert} \chi_n^\Q  (\alpha_0,
\cdots,\alpha_{j-1},\nabla^{End}\alpha_j,\alpha_{j+1},  \cdots  \alpha_{n}
).\\ \end{eqnarray*} \end{cor} {\bf Proof:} The first part of the Corollary 
follows from the  above theorem combined with the fact that \begin{eqnarray*} 
 d\,\left( \chi_n^\Q  (\alpha_0, \cdots, \alpha_{n} )\right)&=& \sum_{j=0}^{n}
(-1)^{\vert \alpha_0\vert +\cdots+ \vert\alpha_{j-1}\vert} \chi_n^\Q 
(\alpha_0, \cdots,\alpha_{j-1}, \nabla^{End}\alpha_j,\alpha_{j+1},  \cdots  \alpha_{n} )\\
&+&   \left[\nabla \,\chi_n^\Q\right]   (\alpha_0, \cdots, \alpha_{n} ).\\
\end{eqnarray*} The last part of the corollary then follows.  \begin{rk} In
particular, this formula yields  that $\chi_n^{\Q}$ is covariantly
constant on $\Omega\left(X, Cl_{<-\frac{{\rm dim}M}{n+1}}(\M,\E)\right)$ which
corresponds to classical pseudodifferential valued forms with values in
operators that lie in the Schatten class ${\cal I}_{n+1}\left(L^2(\M,
\E))\right)$ where \hfill \break \noindent $\left(L^2(\M, \E))\right)\to X$ is the smooth  fibration 
with fibre above $x\in X$ given by $L^2(M_x, E_{\vert_{M_x}})$.  \end{rk}

\section{Chern-Weil type weighted trace cochains }
We continue in the same geometric setting as in the previous section keeping the same notations. 
\begin{thm}
Let   $f_i, i=0, \cdots n$ be  polynomial  functions, then  
the Chern-Weil type weighted trace cochain $ \chi_n^\Q   (f_0 (\Omega), \cdots, f_{n}(\Omega) )$ is generally not closed and we have:
\begin{eqnarray*} 
&{}& d  \,\chi_n^\Q    (f_0 (\Omega), \cdots, f_{n}(\Omega) ) \\
&=&
\frac{1}{q}\cdot \sum_{\vert k\vert=0}^{[\vert d\vert\cdot \omega] +{\rm dim } M} \, \frac{  (-1)^{\vert  k\vert +1}\, k!}{(k+1)! }\,
\sum_{j=1}^{n+1 } {\rm res} \left( f_0(\Omega) \wedge \left(f_1(\Omega)\right)^{(k_1)}\wedge\cdots \right.\\
&{}& \left.\wedge \left(f_{j-1}(\Omega)\right)^{(k_{j-1})} \wedge   \left(\nabla^{End}\Q\right)^{(k_j)}\wedge \left(f_{j}(\Omega)\right)^{(k_j+1)}  \cdots\wedge \left(f_{n}(\Omega)\right)^{(k_{n+1})}\wedge \Q^{-\vert k\vert-1}\right)\\
\end{eqnarray*}
where $\vert d\vert= \sum_{i=0}^n d_i$  with  $d_i$ the degree of the polynomial $f_i$, and  $ \omega$ is the order of the operator valued $2$-form $\Omega$. 
\end{thm}
\begin{rk}
\begin{itemize}
\item
For $n=0$, the weighted Chern-Weil type forms investigated in \cite{PR} read:
$${\rm tr}^\Q(f(\Omega))=\chi_0^\Q(f(\Omega)).$$ The obstruction to their closedness which follows from the above theorem  is given by:
$$d\, {\rm tr}^\Q(f(\Omega))= 
\frac{1}{q}\cdot \sum_{k =0}^{[\vert d\vert\, \omega] +{\rm dim}M } \, 
 \frac{(-1)^{  k +1}\, k!}{k+1}  {\rm res} \left( f_0(\Omega) \wedge   \left(\nabla^{End}\Q\right)^{(k)} \wedge  \Q^{- k-1}\right)$$
which compares with the more compact but maybe less tractable formula \cite{PR}:
$$d\, \chi_0^\Q(f(\Omega))=-\frac{1}{q} {\rm res} \left(f_0(\Omega) \wedge \nabla^{End}\log \Q\right).$$
\item 
As already observed in \cite{PR}  in the case of ordinary weighted Chern-Weil type forms, the above theorem tells us that the more negative  $\omega\cdot \vert d\vert $ becomes,  the  fewer will be    the terms that obstruct the closedness. We come back to this below. 
\end{itemize}
 \end{rk}
{\bf Proof:} 
Differentiating $\chi_n^\Q    (f_0 (\Omega), \cdots, f_{n}(\Omega) )$ yields 
 by Corollary  \ref{cor:dchinQalpha}:
\begin{eqnarray*}
&{}&d \, \chi_n^\Q    (f_0 (\Omega), \cdots, f_{n}(\Omega) )\\
&=&
\frac{1}{q}\cdot \sum_{\vert k\vert=0}^{[\vert a\vert] +{\rm dim} M} \,\frac{k!}{(k+1)!}
\sum_{j=1}^{n+1 } (-1)^{\vert  k\vert +1}  {\rm res} \left( f_0 (\Omega)\wedge  \left(f_{1}(\Omega)\right)^{(k_1)}\wedge  \cdots \wedge \left(f_{j-1}(\Omega)\right)^{(k_{j-1})} \right.\\
&{}& \left. \wedge   \left(\nabla^{End}\Q\right)^{(k_j)}\wedge  \left(f_{j}(\Omega)\right)^{(k_j+1)}\wedge   \cdots\right.\\
&{}&\left.\cdots\wedge   \left(f_{n}(\Omega)\right)^{(k_{n+1})}\wedge \Q^{-\vert k\vert-1}\right)\\
\end{eqnarray*}
where we have used the Bianchi identity $\nabla^{End} \Omega=0$ to 
 cancel the terms \hfill \break \noindent $\chi_n^\Q    \left(f_0 (\Omega), \cdots,f_{j-1}(\Omega)\right)$ and $\nabla^{End} f_j(\Omega),f_{j+1}(\Omega),  \cdots f_{n}(\Omega) ).$
\begin{cor} Let   $\Omega$ have  integer  order $\omega<0$, then any
polynomials  $f_i, i=0, \cdots n$ of orders $d_i, i=0, \cdots, n$ large enough
so  that  $ \vert d\vert >-\frac{{\rm dim} \, M}{\omega}$ give rise to closed 
Chern-Weil type weighted trace cochains $ \chi_n^\Q   (f_0 (\Omega), \cdots,
f_{n}(\Omega) )$.  Moreover, if $\Q_t,\,  t\in ]0, 1[$ is a smooth  one
parameter families of weights with constant order $q$  and $\nabla_t,\,  t\in
]0, 1[$ a smooth family  of connections, the curvatures of which have constant
order $\omega$, then  $$ \frac{d}{dt} \chi_n^{\Q_t}   (f_0 (\Omega_t), \cdots,
f_{n}(\Omega_t) )=  d\, \sum_{j=1}^n\, \sum_{i=0}^{\vert d_j\vert}\,  {\rm
tr}^{\Q_t}\left(\Omega_t^{d_0} \wedge \Omega_t^{d_1} \wedge \cdots\wedge \dot
\nabla_t\wedge \cdots \wedge \Omega_t^{d_n} \wedge  \Q_t^{-\vert k\vert}\right),$$
where $\dot \nabla_t$ stands in the $j$-th position. It follows that  
Chern-Weil type weighted trace cochains $ \chi_n^\Q   (f_0 (\Omega), \cdots,
f_{n}(\Omega) )$ define topological characteristic classes. \end{cor} {\bf
Proof:} The first part of the corollary follows from the above theorem with
$\vert a\vert = \vert d\vert \cdot\omega$. If    $ \vert d\vert >-\frac{{\rm
dim}\, M}{\omega}$, we have $[\vert a\vert] = \vert d\vert\cdot  \omega+{\rm dim}\,
M<0$ so that the sum  involving the residues vanishes. This implies that $d \left(
\chi_n^\Q    (f_0 (\Omega), \cdots, f_{n}(\Omega) )\right)=0$.\\ Let now   $\Q_t,\, 
t\in ]0, 1[$ be a smooth  one parameter families of weights with constant
order $q$  and $\nabla_t,\,  t\in ]0, 1[$ a smooth family  of connections, the
curvatures of which have constant order $\omega$. Replacing $\nabla$ by
$\frac{d}{dt}$ in  Theorem \ref{thm:nablachiQ}, we first  find  that  $
\frac{d}{dt}\bar \chi_{n, \Q_t}(z)$  is holomorphic at zero and for any
$\alpha_0, \cdots, \alpha_n \in \Omega(X, Cl(\M, \E))$ 
\begin{eqnarray*} 
&{}&\left( \frac{d}{dt} \chi_{n}^{ \Q_t}\right)   (\alpha_0, \cdots,
\alpha_{n} )\\ &=&\lim_{z\to 0}\left( \frac{d}{dt} \,\bar \chi_{n,
\Q_t}(z)\right) (\alpha_0, \cdots, \alpha_{n})\\ &=&  \frac{1}{q}\cdot
\sum_{\vert k\vert=0}^{[\vert a\vert] +{\rm dim}\, M} \, \frac{k!}{(k+1)!}
\sum_{j=1}^{n+1 } (-1)^{\vert \alpha_{0}\vert +\cdots +\vert 
\alpha_{j-1}\vert+\vert  k\vert +1}  {\rm res}
\left( \alpha_0 \wedge \alpha_1^{(k_1)}\wedge \cdots\right.\\ &{}& \left.
\cdots\wedge  \,  \alpha_{j-1}^{(k_{j-1})} \wedge    \left(\frac{d}{dt}
\Q_t\right)^{(k_j)}  \cdots\wedge  \alpha_n^{(k_{n+1})} \wedge\Q_t^{-\vert
k\vert-1}\right).\\ 
\end{eqnarray*} 
Setting $\alpha_i= f_i(\Omega_t)$, we have
$ a_i= d_i \cdot \omega$ and  this  tells us that  provided $[\vert a\vert]=
[\vert d\vert \cdot \omega] <- {\rm dim} \, M, $
\begin{equation}\label{eq:tdep}   \left( \frac{d}{dt} \chi_{n}^{ \Q_t}\right) 
 (f_0(\Omega_t), \cdots, f_{n}(\Omega_t) )=0. \end{equation}  Combining  
equation (\ref{eq:tdep}) and  Corollary \ref{cor:dchinQalpha} then  yields 
provided $[\vert a\vert]= [\vert d\vert \cdot \omega] <- {\rm dim} \, M$
(here as before, $\alpha_i= f_i(\Omega_t)$),
 \begin{eqnarray*}  
&{}&\frac{d}{dt}\,\left( \chi_n^{\Q_t}  (\alpha_0, \cdots, \alpha_{n} )\right)\\ &=&\left(\frac{d}{dt}\, \chi_n^{\Q_t}\right)  \left(\alpha_0, \cdots,
\alpha_{n} \right)+ \sum_{j=0}^n \chi_n^{\Q_t}  (\alpha_0,
\cdots,\frac{d}{dt} \alpha_j,\alpha_{j+1},  \cdots  \alpha_{n} )\\ &=&
\sum_{j=0}^n {\rm tr}^{\Q_t}\left(\alpha_0 \wedge \alpha_1 \wedge \cdots\wedge
\frac{d}{dt}\alpha_j\wedge \cdots \wedge \alpha_n \wedge  \Q_t^{-\vert
k\vert}\right)\\ &+&\frac{1}{q}\sum_{j=0}^n (-1)^{\vert \alpha_0\vert +\cdots+
\vert\alpha_{j-1}\vert+\vert k\vert +1\vert} \sum_{\vert k\vert=1}^{[\vert a\vert] +{\rm dim} \, M}
\frac{k!}{(k+1)!}\,
{\rm res} \left(\alpha_0\wedge  \alpha_1^{(k_1)}\wedge\cdots\right. \\ &{ }&
\left.\cdots \wedge \left(\frac{d}{dt} \alpha_j\right)^{(k_j)}\wedge \alpha_j^{(k_j+1)}\cdots \wedge
\alpha_n^{(k_n)} \wedge \Q_t^{-\vert k\vert}\right).\\
 \end{eqnarray*} 
 We saw
that the first term $\left(\frac{d}{dt}\,
\chi_n^{\Q_t}\right)  \left(\alpha_0, \cdots, \alpha_{n} )\right)$ in the first equation   vanishes 
provided $[\vert a\vert]= [\vert d\vert \cdot \omega] <- {\rm dim} \, M.$.\\ 
We claim that the second term in the last equation  also vanishes for similar
reasons in that case. We are therefore left with the first  term in the last
equation, namely  $\sum_{j=0}^{{\rm dim} \, M} {\rm tr}^{\Q_t}\left(\alpha_0 \wedge \alpha_1 \wedge
\cdots\wedge \frac{d}{dt}\alpha_j\wedge \cdots \wedge \alpha_n \wedge 
\Q_t^{-\vert k\vert}\right).$ Since the $f_i$'s are polynomials, by linearity
we can assume that they are of the form $f_i(X)= X^{d_i}$, in  which case we
have: 
\begin{eqnarray*} \frac{d}{dt}\alpha_j&=& \frac{d}{dt}f_j(\Omega_t)\\
&=& \sum_{i=1}^{d_j}\Omega_t^{i-1}\wedge
\left(\frac{d}{dt}\Omega_t\right)\wedge  \Omega_t^{d_j-i-1}\\ &=&
\sum_{i=1}^{d_j}\Omega_t^{i-1}\wedge \left(\nabla_t \, \dot \nabla_t-\dot 
\nabla_t \nabla_t\right)\wedge  \Omega_t^{d_j-i-1}\\ &=&
\sum_{i=1}^{d_j}\Omega_t^{i-1}\wedge \left(\nabla_t^{End} \, \dot
\nabla_t\right)\wedge  \Omega_t^{d_j-i-1}\\ &=&
\sum_{i=1}^{d_j}\nabla_t^{End}\left( \Omega_t^{i-1}\wedge  \, \dot
\nabla_t\wedge  \Omega_t^{d_j-i-1}\right),\\
 \end{eqnarray*} where we have
used Bianchi identity $\nabla_t^{End}\Omega_t=0$ to "pull out" $\nabla^{End}$. 
Since we know by \label{cor:dchinQ} that $\nabla_t$ "commutes" with 
$\chi_n^{\Q_t}$ provided $[a]<-{\rm dim}\, M$,  whenever $\frac{{\rm dim}\,
M}{d}>-\omega$, we can write:  \begin{eqnarray*} &{}& {\rm
tr}^{\Q_t}\left(\alpha_0 \wedge \alpha_1 \wedge \cdots\wedge
\frac{d}{dt}\alpha_j\wedge \alpha_{j+1}\wedge  \cdots \wedge \alpha_n \wedge  \Q_t^{-\vert
k\vert}\right)\\ &{}& \sum_{i=0}^{ d_j} {\rm tr}^{\Q_t}\left(\nabla_t^{End}
\left(\Omega_t^{d_0} \wedge \Omega_t^{d_1} \wedge \cdots\wedge \dot \nabla_t\wedge
\cdots \wedge \Omega_t^{d_n} \wedge  \Q_t^{-\vert k\vert}\right)\right)\\  &=&
d\,  \sum_{i=0}^{\vert d_j\vert} {\rm tr}^{\Q_t}\left(\Omega_t^{d_0} \wedge
\Omega_t^{d_1} \wedge \cdots\wedge \dot \nabla_t\wedge \cdots \wedge \Omega_t^{d_n}
\wedge  \Q_t^{-\vert k\vert}\right),\\  \end{eqnarray*} where $ \dot \nabla_t$
is in the $j$-th position.  Since this term is exact, summing over $j$ yields
the exactness of  $\sum_{j=0}^n \chi_n^{\Q_t}  (\alpha_0, \cdots,\frac{d}{dt}
\alpha_j,\alpha_{j+1},  \cdots  \alpha_{n} )$   and   ends the proof. \\ \\ We
now specialize to a Quillen-Bismut superconnection setting. Let $D\in
C^\infty(X, {\cal A})$ be a smooth (possibly  odd) section of self-adjoint
elliptic pseudodifferential operators  with constant order. A superconnection
$\nabla$  on ${\cal E}$ induces another superconnection $\A:= \nabla+ \gamma
D$ (where $\gamma=1$ in the $\Z_2$-graded case and $\gamma^2=1$ in the non
graded case). Its curvature is  a two form $\A^2$ with values in ${\cal A}$ so
that   at any  point $x\in X$, $\A_x^2$ is a positive elliptic
pseudodifferential operator of order $2$ valued two form. Taking  $\Q:= \A^2$
gives rise to covariantly constant weighted trace cochains:   \begin{thm} The 
$\A^2$-weighted trace cochains $\chi_n^{\A^2}$ are covariantly constant:
$$\left(\A \chi_{n}^{\A^2} \right)   (\alpha_0, \cdots, \alpha_{n} )=0 \quad
\forall \alpha_0, \cdots, \alpha_n \in \Omega(X, Cl(\M, \E)).$$ Also, for any
polynomial functions $f_0, \cdots, f_n$, the  $ \chi_{n}^{\A^2}  
\left(f_0(\A^2), \cdots, f_n(\A^2)\right)$ define closed characteristic
classes which are independent of the choice of  connection  $\nabla$ from
which $\A $ is defined.  \end{thm} {\bf Proof:} This follows from the above
theorem combined with the fact that $\A^{End} \Q= [\A, \Q]=0$ when $\Q=\A^2$. 
\begin{rk} The second part of the theorem is proven in \cite{PS} for
$n=0$. The generalization $n>0$ does not bring anything new since 
clearly, the expression \hfill \break \noindent  $
\chi_{n}^{\A^2}   \left(f_0(\A^2), \cdots, f_n(\A^2)\right)$ is a   linear
combination of terms of the type ${\rm tr}^{\A^2}(f(\A^2))=  \chi_{0}^{\A^2}  
\left(f(\A^2)\right)$. \end{rk} \section*{Appendix A: Relation to  Higson's
cochain  $(A_0, \cdots, A_n) \mapsto \langle A_0, \cdots, A_n\rangle_z$ } The
methods used in this paper to compute the various anomalies/discrepancies, are
somewhat similar in spirit to  methods used in \cite{CM} and \cite{H} in the
computation of a local representative of the Chern character. This appendix
points out to some of the relations. N. Higson introduces in  \cite{H} formula
(4.1) a  multilinear form $(A_0, \cdots, A_n) \mapsto \langle A_0, \cdots,
A_n\rangle_z$  which   relates to $\bar\chi_{n, Q}$ as follows: \begin{prop}
\begin{eqnarray*}  \langle A_0, \cdots, A_n\rangle_z &:=& (-1)^n
\frac{\Gamma(z)}{2\pi i} \int \l^{-z} d\lambda \,  {\rm tr} \left(
A^0(\lambda- Q)^{-1}A_1 (\lambda-Q)^{-1} \cdots A_n (\lambda- Q)^{-1}\right)\\
&=& \Gamma(z+n) \cdot \bar \chi_{n, Q} (z+n) (A_0, \cdots, A_n)\\ &\simeq&
\Gamma(z+n) \,{\rm TR} \left( A_0 \,A_1 \cdots A_n\, Q^{-n-z}\right)\\ &+& 
\sum_{\vert k\vert=1}^{\infty} \frac{(-1)^{\vert k\vert}}{ (k+1)!} 
\Gamma(z+n+\vert k\vert)  \, {\rm TR} \left( A_0 \,A_1^{(k_1)} \cdots
A_n^{(k_n)} Q^{-n-\vert k\vert -z}\right).\\ \end{eqnarray*} \end{prop}
\begin{rk} This last expression compares with   \begin{eqnarray*}  \langle
A_0, \cdots, A_n\rangle_z &\simeq &  \sum_{\vert k\vert=0}^{\infty}
\frac{(-1)^{\vert k\vert}}{ (\vert k\vert +n)!} \, c(k)  \, {\rm TR} \left(
A_0 \,A_1^{(k_1)} \cdots A_n^{(k_n)} \,Q^{-n-\vert k\vert -z}\right)\\
\end{eqnarray*} derived in \cite{H}, in the proof of Proposition 4.14.
\end{rk} {\bf Proof:}  By Lemma A.2 in \cite{H}  we have: $$\tilde \chi_{n,
Q}(t) (A_0, \cdots, A_n)=  \frac{(-1)^n }{t^n\, 2\pi i} \int d\lambda \,
e^{-t\l}\,    {\rm tr} \left( A_0(\lambda- Q)^{-1}A_1 (\lambda-Q)^{-1} \cdots
A_n (\lambda- Q)^{-1}\right).$$ The result then follows from the fact that
$\bar \chi_{n,Q} (z)$ is the Mellin transform of $\tilde \chi_{n, Q}$:  
\begin{eqnarray*}  &{}&\bar \chi_{n, Q}(z) (A_0, \cdots, A_n)\\ &=&
\frac{1}{\Gamma(z)} \int_0^\infty t^{z-1} \tilde \chi_{n, Q}(t) (A_0, \cdots,
A_n)\\ &= & \frac{(-1)^n}{2\pi i\,\Gamma(z)} \int_0^\infty dt\,  t^{-n+z-1}  
\int d\lambda \, e^{-t\l}\,    {\rm tr} \left( A_0(\lambda- Q)^{-1}A_1
(\lambda-Q)^{-1} \cdots A_n (\lambda- Q)^{-1}\right)\\ &=&  \frac{(-1)^n}{2\pi
i} \frac{\Gamma(z-n)}{\Gamma(z)} \int d\lambda \, \l^{n-z}   {\rm tr} \left(
A_0(\lambda- Q)^{-1}A_1 (\lambda-Q)^{-1} \cdots A_n (\lambda- Q)^{-1}\right)\\
&=&   \frac{1}{\Gamma(z)}\langle A_0, \cdots, A_n\rangle_{z -n}\\
\end{eqnarray*} Using formula (\ref{eq:dbarchinq}) then yields the last
identity in the  proposition.  \section*{Appendix B: Proof of formula
(\ref{eq:basicformula})} For any $A, Q\in Cl(M, E)$ such that $Q$ has scalar
top order symbol,  the following holds \cite{Le}(Lemma 4.2): \begin{lem} If
$p, \e, N>0$ satisfy $\frac{N-a}{q}-p-\e<0$ then $$e^{-t Q} A=
\sum_{j=0}^{N-1} \frac{(-t)^j}{j!} A^{(j)} e^{-tQ} + R_N (A, Q, t)$$ where,
for any $c>0$ such that $Q+c$ is invertible, there exists $C>0$ such that
$\Vert R_N(A, Q, t)\, (Q+c)^p\Vert \leq C\, t^{\frac{N-a}{q}-p-\e  }.$
\end{lem} \begin{rk} Writing this for short: $$e^{-t Q} \, A\simeq 
\sum_{j=0}^{\infty} \frac{(-t)^j}{j!} A^{(j)}\, e^{-tQ},$$ and taking a Mellin
transform $Q^{-z} =\frac{1}{\Gamma(z)} \int_0^\infty t^{z-1} e^{-t Q} dt$, we
find (compare with \cite{H} Lemma 4.30): $$Q^{-z}\,  A\simeq 
\sum_{j=0}^{\infty}C_{-z}^j A^{(j)}  \, Q^{-z-j}.$$ This can also be derived
from (compare with \cite{H} Lemma 4.20): $$(\lambda-Q)^{-1} \, A\simeq 
\sum_{j=0}^{\infty} A^{(j)}  \, (\lambda-Q)^{-(j+1)}$$ using a Cauchy formula
$C_z^p L^{z-p}= \frac{1}{ 2\pi i} \int \mu^z(\mu-L)^{-p-1} d\mu$ applied to
$L=\lambda- Q$.  \end{rk} Iterating the above lemma yields \cite{PR}
Proposition B.3: \begin{prop} Given any $A_0, \cdots, A_n\in Cl(M, E)$, for
any $J\in \N$, there exist positive integers $N_1, \cdots, N_n$ such that for
$t>0$ \begin{eqnarray*} &{}& \tilde \chi_{n, Q}(t)(A_0, \cdots, A_n)\\ &=&
\sum_{k_1=0}^{N_1-1}\cdots \sum_{k_n=0}^{N_n-1} \frac{ (-t)^{\vert
k\vert}}{(k+1)!} {\rm tr} \left(A_0 A_1^{(k_1)}\cdots A_n^{(k_n)}
e^{-tQ}\right)+ o(t^J). \end{eqnarray*} \end{prop} This slightly differs from
the statement of \cite{PR} Proposition B.3 where the rest term is an $o(t)$.
But it can easily be seen from \cite{PR} Lemma B.2 that the integers $N_1,
\cdots, N_n$ can  in fact be chosen large enough so that the rest term is  an
$o(t^J)$. \\   Also in \cite{PR} proposition B.3, the formula involves a
constant $$C_k = \int_{\Delta_n}u_0^{k_1}(u_0+u_1)^{k_2} \cdots (u_0+\cdots
+u_{n-1})^{k_n}\, du_0\cdots du_n $$ where $\Delta_n=\{(u_0, \cdots, u_n),
0\leq u_i\leq 1,\, \sum_{i=0}^n u_i=1\}$ is the $n$-th simplex. But setting $
u_0+\cdots +u_{n-1}= 1-u_n$ and integrating over $u_n$ yields $C_k=
\frac{1}{k_n+1} \int_{\Delta_n}u_0^{k_1}(u_0+u_1)^{k_2} \cdots (u_0+\cdots
+u_{n-2})^{k_{n-1}}\, du_0\cdots du_n$ which by induction shows that $$C_k=
\frac{1}{(k_n+1)\, (k_{n-1}+1)\cdots (k_0+1)}= \frac{k!}{(k+1)!}$$ where we
have set  $(k+1)!= (k_n+1)! (k_{n-1}+1)!\cdots (k_0+1)!$ and $k!= k_n!\cdots
k_0!$.   \bibliographystyle{plain} 

\begin{thebibliography}{99}  \bibitem[B]{B}
J.-M. Bismut, \otherterm{The Atiyah-Singer index theorem for families of Dirac
operators: two equation proofs}, Inv. Math. {\bf 83}, p.91-151 (1986)
\bibitem[BGV]{BGV}N. Berline, E. Getzler, M. Vergne, {\bf Heat kernels and
Dirac operators}, Springer-Verlag, 1992 \bibitem[CDP]{CDP} A. Cardona, C.
Ducourtioux,  S. Paycha, {\it From  tracial anomalies to  anomalies in quantum
field theory},  Comm. Math. Phys. {\bf 242}, 31--65 (2003) \bibitem[CDMP]{CDMP} A. Cardona,
C. Ducourtioux, J-P. Magnot, S. Paycha, {\it Weighted traces on algebras of
pseudo-differential operators and geometry on loop groups}, Infinite Dim.
Anal. Quant. Prob. Rel Top. {\bf 5}, No. 4  (2002) 503--540 \bibitem[CM]{CM}
A. Connes, H. Moscovici, {\it The local index formula in non commutative
geometry}, Geom. Funct. Anal. {\bf 5} (2) 174-- 243 (1995) \bibitem[F]{F} D.
Freed, {\it The geometry of loop groups}, Journ.  Diff. Geom. {\bf 28}
223--276 (1988)  \bibitem[G-BVF]{G-BVF} J.M. Gracia Bondia, J.C. Varilly, H.
Figueroa, {\bf Elements of noncommutative geometry}, Birkh\"auser Advanced
Texts, Boston MA (2001) \bibitem[H]{H} N. Higson, \otherterm{The local index
formula in non commutative geometry}, Preprint 2004 \bibitem[JLO]{JLO} A.
Jaffe, A. Lesniewski, K. Osterwalder, {\it Quantum $K$-theory. The Chern
character}, Comm. Math. Phys. {\bf 118} 1--14 (1988) \bibitem[KV]{KV} M.
Kontsevich, S. Vishik, {\it Determinants of elliptic pseudo-differential
operators}, Max Planck Institut preprint, 1994 \bibitem[Le]{Le} M. Lesch, {\it
On the non commutative residue for pseudo-differential operators with
log-polyhomogeneous symbols}, Annals of Global Analysis and Geometry, {\bf 17}
 (1999)  151--187 \bibitem[MN]{MN} R. Melrose, V. Nistor, {\it Homology of
pseudo-differential operators I. Manifolds with boundary}, funct-an/9606005,
june 1999 \bibitem[P]{P} S. Paycha, {\it Renormalized traces as a looking
glass into infinite dimensional geometry}, Inf. Dim. Anal. Quant.Prob. Rel.
Top.,  {\bf 4}, N.2,  p.221-266 (2001)  \bibitem[PR]{PR} S. Paycha, S.
Rosenberg, {\it Curvature on determinant bundles and first Chern forms},
Journ. of Geom. Phys.  {\bf 45}, p. 393--429 (2003) \bibitem[PS]{PS} S.
Paycha, S. Scott, in preparation \bibitem[Q]{Q} D. Quillen,
\otherterm{Superconnections  and the Chern character}, Topology  {\bf 24}
p.89-95  (1985) \bibitem[Sc]{Sc} S. Scott, \otherterm{Zeta-Chern forms and the
local family index theorem}, Preprint 2003 \bibitem[Wo]{Wo} M.
Wodzicki,\otherterm{ Non commutative residue} in Lecture  Notes in Math. {\bf
1283}, Springer Verlag 1987

\end{thebibliography}
\end{document}